\documentclass[a4paper,11pt]{article}
\usepackage{subfloat}
\usepackage{jheppub} 
\usepackage{lineno}
\usepackage{booktabs}
\usepackage{multirow}
\usepackage{array}
\newcolumntype{N}{>{\centering\arraybackslash}m{.5in}}

\newcommand{\orcid}[1]{\href{https://orcid.org/#1}{\includegraphics[width=10pt]{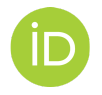}}}

\title{Hamiltonian formulation for scalar model of spontaneous spacetime symmetry violation in gravity}

\author{Jo\~{a}o Victor V. Santos\orcid{0009-0006-3359-3756}}
\author{and Marco Schreck\orcid{0000-0001-6585-4144}}
\affiliation{Departamento de F\'{i}sica, Universidade Federal do Maranh\~{a}o, \\
Campus Universit\'{a}rio do Bacanga, S\~{a}o Lu\'{i}s (MA), 65085-580, Brazil}

\emailAdd{joao.victor2@discente.ufma.br}
\emailAdd{marco.schreck@ufma.br}

\abstract{The focus of this article is on a modification of General Relativity (GR) governed by a dynamical scalar field. The latter is able to acquire a nonzero spacetime-dependent vacuum expectation value, which gives rise to a spontaneous violation of spacetime symmetries. Based on the $(3+1)$ decomposition, we demonstrate how to develop the Hamiltonian formulation for this model. Having done so, our primary interest is to understand how spontaneous spacetime symmetry violation manifests itself in such a setting. In particular, we find that the constraint structure of GR is preserved, although the constraints are clearly modified by the scalar background field. These results emphasize the beauty of spontaneous spacetime symmetry violation in gravity from the viewpoint of the Hamiltonian formulation. They may pose the base for further studies of more sophisticated models of vector and higher-rank tensor fields. Moreover, the description developed can bear fruits when applied within phenomenological quests for spacetime symmetry violation in gravity, in particular, at cosmological scales.}


\begin{document}
\maketitle
\flushbottom

\section{Introduction}
\label{sec:introduction}

General Relativity (GR) has proven to be an astoundingly robust theory of gravity by passing a vast number of experimental tests carried out over the last 100 years~\cite{Einstein:1915bz,Dyson:1920cwa,Pound:1960zz,Everitt:2011hp,Will:2014kxa,Ciufolini:2019ezb,LIGOScientific:2016aoc,LIGOScientific:2018mvr,LIGOScientific:2020ibl,KAGRA:2021vkt,Shao:2021iax,EventHorizonTelescope:2019dse,EventHorizonTelescope:2022xnr}. The geometrization of gravity is highly appealing both from a physical and a mathematical viewpoint and emphasizes how different gravity is compared to the strong and electroweak interactions. Despite the immense success and esthetics of GR, there is a certain chance of it not being the ultimate description of gravitational phenomena at all length scales. On the one hand, at cosmological scales, the $\Lambda$CDM model, which is based on GR, requires a mysterious entity commonly known as Dark Energy \cite{Turner:1998mg} to agree with observational data. Nobody knows anything about the nature of Dark Energy except that it must have a negative pressure. So its need of being introduced into cosmological models could simply be a mere consequence of our lack of knowledge about possibly modified gravitational laws at cosmological scales. On the other hand, the majority of physicists believe that gravity enters a quantum regime at the Planck length, which would modify it in a substantial way. For example, loop quantum gravity~\cite{Thiemann:2007zz} is one approach that tries to construct a consistent unification of quantum theory and gravity.

Some theorists may be motivated to develop a completely new gravity theory, which tackles the aforementioned issues and is most presumably based on quite different principles and mathematics than is GR. A less radial approach to studying hypothetical effects beyond GR at cosmological or microscopic scales is to incorporate modifications into GR at the level of effective field theory. The Einstein-Hilbert (EH) action can serve as a foundation for this endeavor. It is possible to extend this action by contributions that respect certain principles considered to be essential. Other properties inherent to GR such as its basic symmetries or the particular functional dependence of the EH action on the Ricci scalar could be abandoned depending on a researcher's personal taste and interest. Talking about symmetry breaking, general coordinate invariance could be maintained as a reasonable property for a modified-gravity theory to have. However, one may be tempted to include novel terms that either break diffeomorphism symmetry or local Lorentz invariance or both.

A remarkable feature of GR is its symmetry under active diffeomorphisms \cite{Gaul:1999ys}, which is the counterpart to general coordinate invariance. It means that the form of the gravitational laws of nature does not change under a differentiable map of the spacetime manifold onto itself without touching the coordinates at all. In other words, the form of GR remains the same under translations governed by generic vector fields. Local particle Lorentz invariance says that the physical laws in a freely falling lab are locally relativistic with respect to boosts and rotations of an experiment. While in the current paper our interest is on the first of these two concepts, we will also see that they are intimately related to each other.

In general, symmetries can be broken in two manners. The first is explicit breaking, which is a consequence of extending a theory by contributions that are simply devoid of a special symmetry chosen. The modified theory is then no longer invariant under this symmetry. Explicit symmetry violation is known to be plagued by severe issues. For example, an explicit breaking of a gauge symmetry in the Standard Model of elementary particles implies unitarity violations. Nevertheless, explicit symmetry violation can be interesting in its own right. In particular, explicit Lorentz violation in Minkowski spacetime, as it is parameterized by the nongravitational Standard-Model Extension (SME) \cite{Colladay:1996iz,Colladay:1998fq,Kostelecky:2000mm}, does not necessarily lead to any kind of inconsistencies. This definitely constitutes one of the reasons why the SME turned out to be a very successful tool in the search for possible deviations of the laws of high-energy physics from \textit{CPT} and Lorentz invariance \cite{Kostelecky:2008bfz}.

The second way of how to break a symmetry is spontaneously. Spontaneous symmetry violation is a subordinate concept that plays a significant role in a multitude of effects in presumably all areas of physics ranging from condensed-matter to high-energy phenomena. Most crucially, it preserves a certain symmetry of a physical system, but a mechanism gives rise to suitable conditions such that a specific state of the theory, which is commonly the ground state, does not exhibit the symmetry. The most famous realization of this concept in particle physics is the celebrated Higgs mechanism \cite{Higgs:1964pj,Englert:1964et}. The Standard-Model Higgs field takes a nonzero vacuum expectation value, which breaks the electroweak symmetry $\mathit{SU}(2)_L\times \mathit{U}(1)_Y$ down to electromagnetic $\mathit{U}(1)$ invariance. As a consequence, 3 out of the~4 electroweak gauge bosons become massive and Dirac fermions can also acquire masses via suitable Yukawa couplings.

In ref.~\cite{Kostelecky:2003fs} the minimal SME was extended to include gravity. The community quickly realized that an explicit breaking of spacetime symmetries in a curved spacetime manifold imposes involved conditions~\cite{Kostelecky:2003fs,Bluhm:2014oua,Bluhm:2016dzm,Kostelecky:2020hbb} on SME coefficients and spacetime geometry, which are not satisfied in an obvious fashion. While these requirements pose a challenge, one incentive of a series of previous articles~\cite{ONeal-Ault:2020ebv,Reyes:2021cpx,Reyes:2022mvm,Reyes:2023sgk} was to find whether and how inconsistencies of explicit diffeomorphism breaking in the minimal gravitational SME become manifest at a fundamental level within the Hamiltonian formulation \cite{Hanson:1976,Henneaux:1992,Carlip:1998,Bertschinger:2002}. The latter rests upon the $(3+1)$ decomposition of spacetime, which is also known as the ADM formalism~\cite{Arnowitt:1962hi,Arnowitt:2008,Misner:1973}.

To now we have been unable to identify any fundamental inconsistencies that come with explicit symmetry violation. However, the implications of refs.~\cite{Kostelecky:2003fs,Bluhm:2014oua,Bluhm:2016dzm,Kostelecky:2020hbb} are still critical and must be taken into account, as they stand, when explicit violations of spacetime symmetries are considered in gravity. Note that approaches to massive gravity \cite{Duff:1975ik,Rubakov:2004eb,Dubovsky:2004sg}, in particular, dRGT massive gravity \cite{deRham:2010kj,Hassan:2011tf,Kluson:2011rt,deRham:2011ca,Hassan:2011vm,deRham:2011qq,Hassan:2011ea,Kostelecky:2021xhb} explicitly break spacetime symmetries. Nevertheless, such settings were shown to be consistent with Riemannian geometry under certain circumstances \cite{Bluhm:2019ato,Bluhm:2021lzf,Bluhm:2023kph,Reyes:2024ywe}. In more recent times, the phenomenology of explicit spacetime symmetry violation has been explored, in particular, in cosmology \cite{ONeal-Ault:2020ebv,Reyes:2022dil,Khodadi:2023ezj,Reyes:2024hqi} and gravitational waves \cite{Kostelecky:2017zob,Nilsson:2022mzq,Bailey:2023lzy,Bailey:2024zgr}.

Controlling coefficients of the gravitational SME have already been constrained in a broad range of experiments such as gravitational waves \cite{Kostelecky:2016kfm,Mewes:2019dhj,Shao:2020shv,Wang:2021ctl,ONeal-Ault:2021uwu,Niu:2022yhr,Haegel:2022ymk}, gravitational Cherenkov radiation \cite{Moore:2001bv,Kostelecky:2015dpa}, pulsar-timing measurements \cite{Bailey:2006fd,Dong:2023nau,Xu:2020zxs}, and table-top experiments \cite{Bailey:2006fd,Shao:2016cjk,Kostelecky:2021tdf,Ivanov:2021bvk,Zhang:2023hey}; see ref.~\cite{Kostelecky:2008bfz} for a comprehensive compilation of constraints. It is reasonable to distinguish between constraints on coefficients for spontaneous and explicit symmetry violation. In general, the latter are associated with experimental searches for beyond-Riemannian spacetime geometry \cite{Kostelecky:2020hbb,Kostelecky:2021tdf}.

Our interest in the current paper is to apply the machinery developed in refs.~\cite{ONeal-Ault:2020ebv,Reyes:2021cpx,Reyes:2022mvm,Reyes:2023sgk} to a modified-gravity theory with diffeomorphism symmetry broken spontaneously. By doing so, we intend to find how the crucial differences between explicit and spontaneous diffeomorphism violation show up in the Hamiltonian formulation of the theory, in particular, its constraint structure. Our analysis will be carried out at the level of modified GR. Thus, the model will not be linearized in a metric perturbation, whereupon our findings are applicable in the strong-field regime, too. We will be making some generic statements on spontaneous spacetime symmetry violation, which, to our best knowledge, are not to be found anywhere else in the contemporary literature.

The exploration of spontaneous violations of spacetime symmetries has been a successful and popular research program for several decades. In 1963, Bjorken interpreted the photon as a massless Goldstone boson in a theory with spontaneous Lorentz symmetry breaking~\cite{Bjorken:1963vg}. This finding inspired others to take the proposed concept over into a gravitational setting to explain the graviton as a Goldstone boson arising from spontaneous symmetry violation in GR~\cite{Phillips:1966zzc,Ohanian:1970qe}. In fact, spontaneous Lorentz violation was also demonstrated to occur in certain open-string field theories~\cite{Kostelecky:1988zi,Kostelecky:1989jp,Kostelecky:1989jw,Kostelecky:1991ak,Kostelecky:1994rn} being the principal discovery that triggered the construction of the SME.

Research performed after the year 2000 delved into different ideas on spontaneous Lorentz violation in electrodynamics~\cite{Alfaro:2009iv,Escobar:2015gia,Escobar:2018oun} as well as spontaneous symmetry violation in gravity~\cite{Kraus:2002sa,Arkani-Hamed:2003pdi,Berezhiani:2008ue}. A particular class of theories widely studied in the literature are known as bumblebee models. The latter usually involve a dynamical vector-valued field in a suitable potential. Such models have been of vast interest both in a nongravitational context~\cite{ArkaniHamed:2004ar,Bluhm:2006im,Bluhm:2007bd,Bluhm:2008yt,Hernaski:2014jsa,Bonder:2015jra,Seifert:2018mmr} and in the presence of gravity~\cite{Eling:2003rd,Carroll:2004ai,Gripaios:2004ms,Bluhm:2004ep,Seifert:2009gi,Colladay:2019lig,Bonder:2021gjo,Amarilo:2023wpn}.

In the absence of gravity, the emergence of a nonzero vector-like vacuum expectation value breaks particle Lorentz invariance \cite{Colladay:1998fq} spontaneously. When gravity is present, the vacuum expectation value 
generically depends on the spacetime coordinates, which gives rise to spontaneous particle diffeomorphism violation \cite{Bluhm:2014oua} as well as local particle Lorentz violation. Bumblebee models have a rather simple form, but they have the potential of exhibiting interesting properties, which makes them an attractive subject of study. For example, an active research direction that rests upon such bumblebee models is the search for modified black-hole solutions~\cite{Casana:2017jkc,Ding:2019mal,Maluf:2020kgf,Gullu:2020qzu,DCarvalho:2021zpf,Poulis:2021nqh,Xu:2022frb,Mai:2023ggs,Xu:2023xqh,Mai:2024lgk}, where this list merely poses an excerpt of a comprehensive compilation of papers. Extensions of bumblebee models that involve antisymmetric second-rank tensor fields~\cite{Altschul:2009ae,Kostelecky:2009zr,Carroll:2009mr,Seifert:2010mj,Seifert:2010uu,Hernaski:2016dyk,Assuncao:2019azw,Seifert:2019kuz} have been considered, too.

The Hamiltonian formulation of bumblebee-type models has not been extensively developed in the literature, so far. Doing a full analysis of the constraint structure of such models is expected to be challenging because of the presence of a vector-valued field. Therefore, a first reasonable step is to explore a modified-gravity theory with symmetry violation parameterized by the scalar field denoted as $u=u(x)$ in SME literature \cite{Kostelecky:2003fs}. Unlike in previous papers on the Hamiltonian formulation of the minimal gravitational SME \cite{ONeal-Ault:2020ebv,Reyes:2021cpx,Reyes:2022mvm,Reyes:2023sgk}, the latter is now taken as a dynamical field, i.e., the modified EH action will be supplemented by additional kinetic and potential terms for $u$.

Note that scalar fields have played a substantial role in high-energy physics for the past decades and their most prominent application is obviously in the Higgs mechanism~\cite{Higgs:1964pj,Englert:1964et}. Moreover, exotic scalar fields are used in descriptions of the inflationary period of our Universe~\cite{Guth:1980zm}, Dark-Matter models based on axions~\cite{Abbott:1982af,ADMX:2018gho}, string theory \cite{Green:2012} as well as in scalar-tensor models of gravity~\cite{Brans:1961sx,Damour:1992we}. The current manuscript makes use of the latter.

Our paper is organized as follows. The model to be studied is introduced and described in section~\ref{sec:models}, accompanied by a brief summary of the generic characteristics of bumblebee-type models. Subsequently, its $(3+1)$ decomposition is carried out in section~\ref{eq:ADM-decomposition}. Section~\ref{sec:potential} provides a description of how a suitably chosen potential can lead to a vacuum expectation value for $u$ and which are its implications on violations of spacetime symmetries. Section~\ref{sec:field-equations} goes into more detail on the dynamics of the theory by discussing the field equations of the gravitational and the scalar fields. The forthcoming section~\ref{sec:hamiltonian-formulations} is dedicated to the development of the Hamiltonian formulation for the model. The modified Hamiltonian and momentum constraints as well as their relationship with generating the relevant spacetime symmetries are elaborated in 
section~\ref{sec:symmetry-generators-field-equations}. In addition, we observe how the constraints are connected to suitable projections of the covariant field equations. The constraint structure is analyzed in section~\ref{sec:constraint-structure}, which allows us to make a statement on the number of propagating gravitational degrees of freedom. Finally, we briefly summarize our findings and provide an outlook in section~\ref{sec:final-remarks}. Calculational details worthwhile to consider but of secondary interest to the reader are relegated to appendices~\ref{sec:field-equations-derivations}, \ref{eq:constraint-algebra-computation}. Natural units with $c=G_N=1$ are employed, unless otherwise stated.

\section{Models of spontaneous diffeomorphism violation}
\label{sec:models}

Here, we would like to introduce the specific model investigated in the remainder of the paper. Furthermore, the essential characteristics of bumblebee models will be discussed, too.

\subsection{Scalar-tensor gravity model}

Let $S$ be the action of a particular scalar-tensor gravity theory \cite{Damour:1992we} supplemented by an extended Gibbons-Hawking-York (GHY) boundary term \cite{York:1972sj,Gibbons:1976ue}:
\begin{align}
\label{eq:action-BD-type-theory}
S_{\mathrm{ST}}&=\int_{\mathcal{M}} \mathrm{d}^4x\,\frac{\sqrt{-g}}{2\kappa}\,\left(\phi{}^{(4)}R-\frac{\omega}{\phi}g^{\mu\nu}\partial_{\mu}\phi\partial_{\nu}\phi-U(\phi)\right)+2\oint_{\partial\mathcal{M}} \mathrm{d}^3y\,\varepsilon\frac{\sqrt{q}}{2\kappa}\,\phi K\,,
\end{align}
where the spacetime manifold $\mathcal{M}$ is parameterized by coordinates $x^{\mu}$. Furthermore, $\kappa=8\pi G_N$ with the Newtonian constant $G_N$ and $g_{\mu\nu}$ is the metric of the spacetime manifold $\mathcal{M}$ with $g:=\det(g_{\mu\nu})$ being its determinant and $g^{\mu\nu}$ its inverse. The intrinsic geometry of $\mathcal{M}$ is characterized by the Ricci scalar ${}^{(4)}R:=g^{\mu\nu}{}^{(4)}R_{\mu\nu}$ obtained from the Ricci tensor ${}^{(4)}R_{\mu\nu}$. The dynamical scalar field is denoted as $\phi$ and the first term of eq.~\eqref{eq:action-BD-type-theory} describes how the latter couples to GR. The second term incorporates the kinematics of the scalar field where $\omega$ is a dimensionless physical parameter, which can be constrained by experimental searches for deviations from GR. The third term contains a suitable potential $U(\phi)$ with a single global minimum or several ones at values $\phi\neq 0$. For now, the potential is kept as a generic function, but the reader may wish to think of a coordinate-dependent double-well potential. Note that the bulk part of eq.~\eqref{eq:action-BD-type-theory} is related to Brans-Dicke theory \cite{Brans:1961sx}; see also ref.~\cite{Reyes:2022mvm}.

To render the principle of stationary action well-defined, we include a term on the spacetime boundary $\partial\mathcal{M}$, which is parameterized by generic coordinates $y^a$ \cite{Reyes:2021cpx,Reyes:2022mvm,Reyes:2023sgk}. Moreover, $q_{ab}$ is the pullback of $g_{\mu\nu}$ onto $\partial\mathcal{M}$, i.e., the metric $q_{ab}$ describes the intrinsic geometry on the spacetime boundary. As before, $q:=\det(q_{ab})$ is the corresponding determinant and we call $K:=q^{ab}K_{ab}$ the trace of the extrinsic-curvature tensor $K_{ab}$, which arises due to the embedding of $\partial\mathcal{M}$ into $\mathcal{M}$. Now, $\varepsilon=n^2$ with the generic unit vector $n^{\mu}$ normal to $\partial\mathcal{M}$. For spacelike (timelike) parts of the boundary it holds that $\varepsilon=\mp 1$. In principle, $\varepsilon=0$ for lightlike parts, but they form a set of measure zero and do not provide contributions to the boundary integral. For a recent more detailed discussion of the physics of extended GHY boundary terms in a setting of explicit diffeomorphism violation the reader may consult ref.~\cite{Reyes:2023sgk} if they so wish.

\subsection{Bumblebee models}

Bumblebee models are relatively simple field theory models describing a spontaneous breakdown of global or local particle Lorentz invariance, depending on the context. In spite of being relatively simple in form, they nevertheless possess a reasonably large explanatory power and encapsulate many of the most interesting aspects of spontaneous particle Lorentz violation. A typical bumblebee model in Minkowski spacetime involves a vector field often denoted as $B_{\mu}$ that has a potential of the form $V(B_{\mu}B^{\mu}-b^2)$ associated with it. A suitable choice can be a positive, quadratic function such as $V(x)=\lambda x^2/2$ with a coupling constant~$\lambda$. Here $b^2=\langle B_{\mu}\rangle\langle B^{\mu}\rangle$ where $\langle\bullet\rangle$ indicates a vacuum expectation value.

Thus, in resemblance to the Higgs mechanism in the Standard Model, the vector field~$B_{\mu}$ can undergo spontaneous symmetry violation such that the ground state of the theory is characterized by a preferred spacetime direction given by the vacuum expectation value $\langle B_{\mu}\rangle$. The field $B_{\mu}$ chooses the latter out of an infinite number of possibilities. The Standard Model does not contain a bumblebee-type field, of course. However, it was shown that in the context of bosonic string field theory, the presence of a tachyon mode is capable of giving rise to potentials for tensor fields such that the latter can naturally exhibit nonvanishing vacuum expectation values \cite{Kostelecky:1988zi}. Such a mechanism is possible, since the gauge structures of string theories are way less restricted than those of ordinary gauge field theories employed in the construction of the Standard Model.

When we turn from Minkowski spacetime to a curved spacetime, the overall behavior is more involved and we must distinguish between particle diffeomorphism symmetry of the spacetime manifold itself and local particle Lorentz invariance in a freely falling inertial frame. Although in the current paper we will not make use of the vierbein formalism in computations, it is highly suitable in the current context to explain the physics. Let $e^a_{\phantom{a}\mu}$ and $e_{\mu}^{\phantom{\mu}a}$ be vierbeins that transform tensor fields from a generic spacetime tangent frame to a freely falling inertial frame and vice versa. Greek indices are then usually used in the former, whereas Latin indices are adequate for the latter. Now, in the presence of gravity, it is natural to take a particular bumblebee model from Minkowski spacetime over to a freely falling inertial frame, where the geometry is Minkowskian. This procedure provides a bumblebee field denoted as $B_a$, which can acquire a vacuum value $\langle B_a\rangle$ such that the emergence of a preferred direction in the local frame leads to a breakdown of local particle Lorentz invariance. The existence of the vierbein then allows us to promote the bumblebee field to a field $B_{\mu}$ living in a generic spacetime tangent frame. By doing so, $B_{\mu}$ gives rise to a violation of diffeomorphism invariance, i.e., both apparently very different types of symmetry violations are closely related to each other \cite{Bluhm:2004ep}.

What we intend to study in the current paper is not a vector-valued field, but we will introduce a scalar field denoted as $u=u(x)$ in the SME literature, whose dynamics is described by a bumblebee-type theory. To do so, we start from the scalar-tensor gravity model of eq.~\eqref{eq:action-BD-type-theory}. Since the model contains a dynamical scalar field $\phi=\phi(x)$ within a potential $U(\phi)$, a suitable choice of the potential will permit that the scalar field acquire a vacuum value $\langle\phi\rangle(x)$. Since the latter is dependent on spacetime position, $\langle\phi\rangle(x)$ defines a scalar background field on the spacetime manifold, i.e., particle diffeomorphism invariance is violated spontaneously. In addition, $\partial_a\langle\phi\rangle(x)$ also provides a preferred vector-valued background field in each local frame, whereupon local particle Lorentz invariance is violated, too. Before we delve into the peculiarities of the model depending on $u$, we will be performing a $(3+1)$ decomposition of the generic eq.~\eqref{eq:action-BD-type-theory}, which shall be our specific tool of choice to understand the physics of $u$.

\section{ADM decomposition}
\label{eq:ADM-decomposition}

The $(3+1)$ decomposition \cite{Arnowitt:1962hi,Arnowitt:2008,Misner:1973,Hanson:1976,Henneaux:1992,Carlip:1998,Bertschinger:2002} of eq.~\eqref{eq:action-BD-type-theory} is performed by foliating $\mathcal{M}$ into purely spacelike hypersurfaces $\Sigma_t$ of constant coordinate time $t$. Then, the spacetime metric $g_{\mu\nu}$ is decomposed into parts $g_{00}$, $g_{0a}$, and $q_{ab}:=g_{ab}$ where $q_{ab}$ contains the physical degrees of freedom of the modified-gravity theory and $g_{00},g_{0a}$ involve gauge degrees of freedom. These are parameterized in terms of the lapse function $N=:1/\sqrt{-g^{00}}$ and the shift vector $N^a$ with $N_a=:g_{0a}$. We then use
\begin{align}
\label{eq:ricci-scalar-decomposition}
\phi{}^{(4)}R=\phi(R+K^2+K_{ab}K^{ab})+\frac{2}{N}\phi\mathcal{L}_mK-\frac{2}{N}\phi D_aD^aN\,,
\end{align}
where $R$ is the Ricci scalar and $D_a$ the $q_{ab}$-compatible covariant derivative on $\Sigma_t$. In addition, $\mathcal{L}_m$ denotes the Lie derivative \cite{Carroll:1997ar,Yano:1957} with respect to the direction $m^{\mu}:=Nn^{\mu}$. Since $\phi$ is a scalar, the Lie derivative is simply a directional derivative along $m^{\mu}$ such that
\begin{equation}
\label{eq:lie-derivative-scalar}
\frac{1}{N}\mathcal{L}_m\phi=n^{\mu}\nabla_{\mu}\phi=\frac{1}{N}(\dot{\phi}-\mathcal{L}_{\mathbf{N}}\phi)\,.
\end{equation}
The dot denotes a time derivative and $\mathcal{L}_{\mathbf{N}}\phi=N^aD_a\phi$ is the Lie derivative of $\phi$ with respect to the shift vector, which is a directional derivative, too. We then reformulate the corresponding term as
\begin{equation}
\frac{\phi}{N}\mathcal{L}_mK=\nabla_{\mu}(n^{\mu}\phi K)-\phi K^2-\frac{1}{N}K\mathcal{L}_m\phi\,.
\end{equation}
The first term on the right-hand side of the latter is a total covariant derivative, which, upon integration, gives rise to a boundary term canceling the GHY-type boundary term in eq.~\eqref{eq:action-BD-type-theory}. This is everything that we need to ADM-decompose the gravitational-coupling term in eq.~\eqref{eq:action-BD-type-theory}.

Next, we dedicate ourselves to the kinematic contribution. Here, the completeness relation
\begin{equation}
g^{\mu\nu}=q^{\mu\nu}-n^{\mu}n^{\nu}\,,
\end{equation}
is valuable and we arrive at:
\begin{equation}
\label{eq:decomposition-kinetic-term}
g^{\mu\nu}\partial_{\mu}\phi\partial_{\nu}\phi=(q^{\mu\nu}-n^{\mu}n^{\nu})\nabla_{\mu}\phi\nabla_{\nu}\phi=q^{\mu\nu}\nabla_{\mu}\phi\nabla_{\nu}\phi-n^{\mu}n^{\nu}\nabla_{\mu}\phi\nabla_{\nu}\phi\,.
\end{equation}
We also employ the decomposition formula for the $g_{\mu\nu}$-covariant derivative of a generic spacetime tensor \cite{Gourgoulhon:2007ue,Gourgoulhon:2012}, which reads
\begin{equation}
D_{\varrho}T^{\alpha_1\dots \alpha_p}_{\phantom{\alpha_1\dots \alpha_p}\beta_1\dots\beta_q}=q^{\alpha_1}_{\phantom{\alpha_1}\mu_1}\dots q^{\alpha_p}_{\phantom{\alpha_p}\mu_p}q^{\nu_1}_{\phantom{\nu_1}\beta_1}\dots q^{\nu_q}_{\phantom{\nu_q}\beta_q}q^{\sigma}_{\phantom{\sigma}\varrho}\nabla_{\sigma}T^{\mu_1\dots\mu_p}_{\phantom{\mu_1\dots\mu_p}\nu_1\dots\nu_q}\,.
\end{equation}
Applying this formula to the first term of eq.~\eqref{eq:decomposition-kinetic-term} implies
\begin{equation}
q^{\mu\nu}\nabla_{\mu}\phi\nabla_{\nu}\phi=q^{\mu}_{\phantom{\mu}\varrho}q^{\nu}_{\phantom{\nu}\sigma}q^{\varrho\sigma}\nabla_{\mu}\phi\nabla_{\nu}\phi=q^{\varrho\sigma}D_{\varrho}\phi D_{\sigma}\phi=q^{ab}D_a\phi D_b\phi\,.
\end{equation}
For the second term in eq.~\eqref{eq:decomposition-kinetic-term} we use eq.~\eqref{eq:lie-derivative-scalar} for the Lie derivative leading to
\begin{equation}
n^{\mu}n^{\nu}\nabla_{\mu}\phi\nabla_{\nu}\phi=\frac{1}{N^2}(\dot{\phi}-N^aD_a\phi)^2\,.
\end{equation}
Finally, the ADM-decomposed action resulting from eq.~\eqref{eq:action-BD-type-theory} is
\begin{subequations}
\label{eq:BD-action-ADM-decomposed}
\begin{align}
S_{\mathrm{ST}}^{(3+1)}&=\int_{t_1}^{t_2}\mathrm{d}t\,\int_{\Sigma_t}\mathrm{d}^3y\,\mathcal{L}_{\mathrm{ST}}^{(3+1)}\,, \displaybreak[0]\\[2ex]
\mathcal{L}_{\mathrm{ST}}^{(3+1)}&=\mathcal{L}_R+\mathcal{L}_{\omega}+\mathcal{L}_U\,, \displaybreak[0]\\[2ex]
\label{eq:lagrangian-R}
\mathcal{L}_R&=\frac{\sqrt{q}}{2\kappa}\Big\{N\phi(R-K^2+K_{ab}K^{ab})-2\Big[K(\dot{\phi}-N^aD_a\phi)+\phi D_aD^aN\Big]\Big\}\,, \displaybreak[0]\\[2ex]
\mathcal{L}_{\omega}&=-\frac{N\sqrt{q}}{2\kappa}\frac{\omega}{\phi}\bigg[q^{ab}D_a\phi D_b\phi-\frac{1}{N^2}(\dot{\phi}-N^aD_a\phi)^2\bigg]\,, \displaybreak[0]\\[2ex]
\mathcal{L}_U&=-\frac{N\sqrt{q}}{2\kappa}U(\phi)\,.
\end{align}
\end{subequations}
At this point we can add a $q_{ab}$-compatible total derivative to the action of the form
\begin{equation}
D_a(\phi D^aN)=D_a\phi D^aN+\phi D_aD^aN\,.
\end{equation}
In principle, the latter gives rise to a boundary term on $\partial\Sigma_t$. Introducing it permits that we replace the double $q_{ab}$-compatible covariant derivative of the lapse function, $-\phi D_aD^aN$, by $q^{ab}D_aND_b\phi$. Within the action, eq.~\eqref{eq:lagrangian-R} can then be recast into
\begin{equation}
\mathcal{L}_R'=\frac{\sqrt{q}}{2\kappa}\Big\{N\phi({}^{(3)}R-K^2+K_{ab}K^{ab})-2K(\dot{\phi}-N^aD_a\phi)+2q^{ab}D_aND_b\phi\Big\}\,.
\end{equation}
After doing so, the ADM decomposition of the scalar-tensor gravity theory of eq.~\eqref{eq:action-BD-type-theory} agrees with the result provided in ref.~\cite{GabrieleGionti:2020drq}.

To construct a model of spontaneous spacetime symmetry violation in the spirit of ref.~\cite{Bluhm:2004ep}, we parameterize the scalar field $\phi$ in terms of the scalar field called $u$ in the gravitational sector of the SME \cite{Kostelecky:2003fs,Kostelecky:2020hbb}: $\phi=1-u$. By doing so, eq.~\eqref{eq:BD-action-ADM-decomposed} takes the following form:
\begin{subequations}
\label{eq:u-action-ADM-decomposed}
\begin{align}
S_u^{(3+1)}&=\int_{t_1}^{t_2}\mathrm{d}t\,\int_{\Sigma_t}\mathrm{d}^3y\,\mathcal{L}_u^{(3+1)}\,, \displaybreak[0]\\[2ex]
\mathcal{L}_u^{(3+1)}&=\mathcal{L}_{R,u}+\mathcal{L}_{\omega,u}+\mathcal{L}_V\,, \displaybreak[0]\\[2ex]
\label{eq:lagrange-density-R-u}
\mathcal{L}_{R,u}&=\frac{\sqrt{q}}{2\kappa}\Big\{N(1-u)(R-K^2+K_{ab}K^{ab}) \notag \\
&\phantom{{}={}}\hspace{0.8cm}+2\Big[K(\dot{u}-N^aD_au)-(1-u)D_aD^aN\Big]\Big\}\,, \displaybreak[0]\\[2ex]
\label{eq:lagrange-density-kinematics-u}
\mathcal{L}_{\omega,u}&=-\frac{N\sqrt{q}}{2\kappa}\frac{\omega}{1-u}\bigg[q^{ab}D_auD_bu-\frac{1}{N^2}(\dot{u}-N^aD_au)^2\bigg]\,, \displaybreak[0]\\[2ex]
\mathcal{L}_V&=-\frac{N\sqrt{q}}{2\kappa}V(u)\,,
\end{align}
\end{subequations}
where we denote $U(1-u)=:V(u)$. Now, the first part of the latter theory corresponds to a modified ADM-decomposed EH Lagrange density; cf.~refs.~\cite{Reyes:2021cpx,Reyes:2022mvm}. Without the kinetic and potential terms, $\mathcal{L}_{\omega,u}$ and $\mathcal{L}_V$, respectively, the scalar field $u$ already acts as a background, which generically depends on the spacetime coordinates. Consequently, particle diffeomorphism symmetry is broken explicitly. Certain aspects of the Hamiltonian formulation of this setting have already been studied in refs.~\cite{Reyes:2021cpx,Reyes:2022mvm,Reyes:2023sgk} with great detail. In the presence of $\mathcal{L}_{\omega,u}$ and $\mathcal{L}_V$, at least some of the previously obtained results are expected to change drastically, in particular, when it comes to the constraint structure of such a theory.

\section{Potential, vacuum expectation value, and spontaneous symmetry breaking}
\label{sec:potential}

What is new in the current approach, when compared to refs.~\cite{Reyes:2021cpx,Reyes:2022mvm,Reyes:2023sgk}, is that kinetic and potential terms render the scalar field $u$ dynamical. The theory based on eq.~\eqref{eq:u-action-ADM-decomposed} then has two thermodynamic phases. The first phase is established before spontaneous spacetime symmetry violation. The scalar field $u$ then satisfies its own field equation and transforms under particle diffeomorphisms as expected. The second phase arises after spontaneous symmetry breaking, i.e., when $u$ evolves to its ground state. The latter would correspond to a field configuration in the global minimum of the potential $V(u)$.  In the case of a degenerate set of global minima, $u$ would pick a single one of these at random.

Although we work in a classical setting, the chosen value of $u$ in the ground state will be referred to as a vacuum expectation value $\langle u\rangle$ with the vacuum state arising from a fundamental quantized theory hitherto unbeknownst to us. The vacuum value $\langle u\rangle$ is fixed under particle diffeomorphisms, whereupon it takes the characteristics of a background field. Hence, by choosing some $\langle u\rangle\neq 0$, eq.~\eqref{eq:u-action-ADM-decomposed} undergoes a spontaneous violation of particle diffeomorphism invariance. The phase transition is expected to occur at a critical temperature some time after the Big Bang, analogously to the Higgs mechanism. However, the typical temperatures for a (hypothetical) spontaneous violation of spacetime symmetries and electroweak gauge symmetry are expected to significantly differ from each other due to the inherently different physics involved.
\begin{figure}
    \centering
    \includegraphics[scale=0.45]{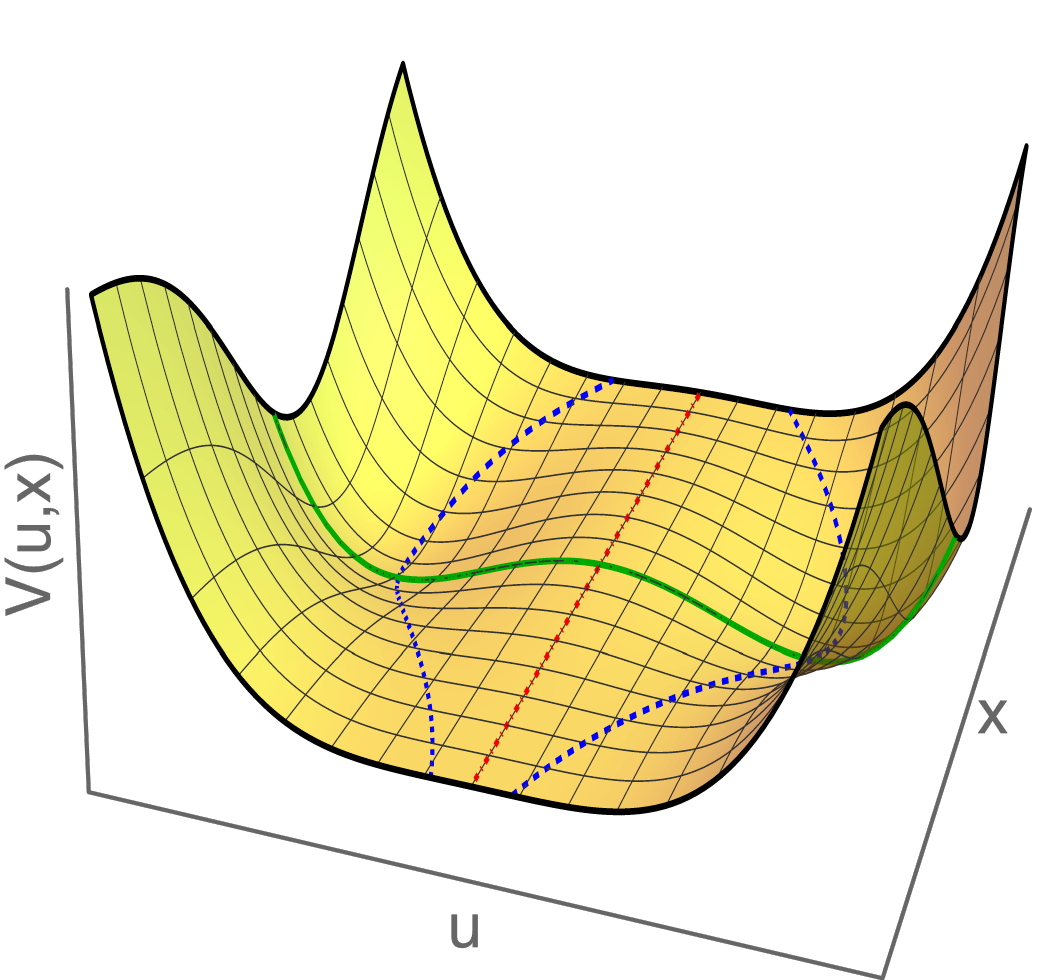}
    \caption{Double-well potential of eq.~\eqref{eq:mexican-hat-potential} as a function of the scalar field $u$ and a single spatial coordinate $x$. The potential for a fixed $x=x_0$, which is of double-well form, is shown as a plain, green curve. The dashed-dotted, red curve represents the set of local maxima and the dashed, blue curves show the global minima as functions of $x$.}
    \label{fig:double-well-potential-potential-space}
\end{figure}

Now, we suitably choose $V(u)$ in the form of a double-well potential:
\begin{equation}
\label{eq:mexican-hat-potential}
V(u)=-\mu^2u^2+\kappa u^4\,,
\end{equation}
where $\mu=\mu(x)\in\mathbb{R}$ and $\kappa=\kappa(x)\geq 0$ are position-dependent functions, so is $u(x)$. For a fixed $x=x_0$, the real scalar field $u$ can acquire one out of three possible vacuum values:
\begin{equation}
\label{eq:vacuum-values-u}
\langle u\rangle(x_0)=\left\{\begin{array}{cl}
\pm\Xi & \text{for } \Xi\in\mathbb{R}\setminus\{0\} \\
0 & \text{for } \Xi=0\,, \\
\end{array}
\right.\quad \Xi=\sqrt{\frac{\mu^2(x_0)}{\kappa(x_0)}}\,.
\end{equation}
Note that this situation differs from that of the Higgs mechanism in various aspects. First, the parameters of the Higgs potential are position-independent. On the contrary, in the current setting one must imagine that there is double-well potential for $u(x)$ with potentially different parameters at each spacetime point. Second, in contrast to $u$, the Higgs field $\Phi$ is not only complex, but also $\mathit{SU}(2)_L\times\mathit{U}(1)_Y$-valued according to the gauge symmetry of the electroweak sector of the Standard Model before electroweak symmetry breaking. Thus, the Higgs field chooses its vacuum value from a continuous, infinite set of possibilities realized in a Mexican-hat potential $V_m(|\Phi|)$ with $|\Phi|=\sqrt{\Phi^{\dagger}\Phi}$. Note that $V_m(|\Phi|)$ has the form of a double-well potential when sliced at a fixed complex $\mathit{U}(1)_Y$ phase. The background field $u$ is real, though, and can choose from 3 different possibilities at each $x$, as described in the following.

\begin{figure}
    \centering
    \includegraphics[scale=0.45]{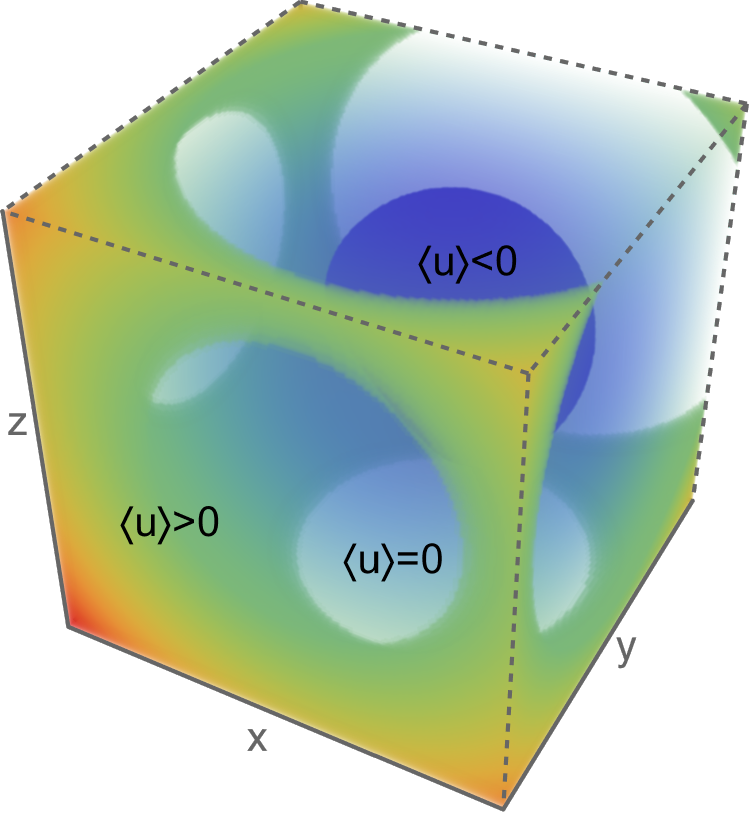}
    \caption{Snapshot of different configurations of $\langle u\rangle$ in three-dimensional space. Different spacetime regions according to eq.~\eqref{eq:regions-u-spacetime} are color-coded.}
    \label{fig:configurations-u}
\end{figure}

Depending on the explicit values of $\mu(x_0)$ and $\kappa(x_0)$, the vacuum value $\langle u\rangle(x_0)$ can be either positive or negative according to eq.~\eqref{eq:vacuum-values-u} or 0. Thus, the potential $V(u)$ gives rise to a nontrivial ground state of $u(x)$ described by a vacuum value $\langle u\rangle(x)$; see figure~\ref{fig:double-well-potential-potential-space}. Assuming that the latter varies smoothly with $x$, three different classes of spacetime regions $R_+$, $R_0$, and $R_-$ are expected such that
\begin{equation}
\label{eq:regions-u-spacetime}
\langle u\rangle(x)\left\{\begin{array}{cl}
>0 & \text{for } x\in R_+ \\
=0 & \text{for } x\in R_0 \\
<0 & \text{for } x\in R_-\,. \\
\end{array}
\right.
\end{equation}
Situations resembling the behavior of the magnetization in certain ferromagnetic materials could also be imagined with domains where $\langle u\rangle(x)>0$ directly neighboring domains with $\langle u\rangle(x)<0$ such that domain walls emerge between both; see figure~\ref{fig:configurations-u} for a visualization. The emergence of topological effects for different types of backgrounds was studied in ref.~\cite{Seifert:2010mj}.

The fields $\mu(x)$ and $\kappa(x)$ of the potential can be taken as vacuum values themselves arising from fundamental physics at the Planck scale such as strings. In contrast, the vacuum value of $u(x)$ emerges dynamically within the effective theory proposed. The detailed forms of $\mu(x)$ and $\kappa(x)$ are not a must-have for the general conclusions that we will draw from our analysis. At the maximum, a number of 10 symmetry generators for diffeomorphisms and local Lorentz transformations can be broken~\cite{Bluhm:2004ep}. It is the form of $\langle u\rangle(x)$ that dictates which ones are broken and preserved, respectively.

There are three types of diffeomorphisms referred to in the literature. First, active diffeomorphisms $x^{\mu}\mapsto x^{\mu}+\xi^{\mu}$ \cite{Gaul:1999ys} can be interpreted as position-dependent translations, i.e., they are generated by vector fields $\xi^{\mu}=\xi^{\mu}(x)$. These diffeomorphisms map each point $p\in\mathcal{M}$ to another point $p'\in\mathcal{M}$ such that the manifold is mapped onto itself. Generically, this also implies that a tensor $T^{\mu\nu\dots}$ in the tangent space at $p$ is mapped to a tensor $T'^{\mu\nu\dots}$ in the tangent space at $p'$. Performing a pull-back operation of $T'^{\mu\nu\dots}$ to $p$, both tensors are found to differ by the Lie derivative of the original tensor along $\xi^{\mu}$, i.e., $\mathcal{L}_{\xi}T^{\mu\nu\dots}$ \cite{Bluhm:2004ep}.

Second, observer/passive diffeomorphisms \cite{Gaul:1999ys,Bluhm:2014oua} are simply general coordinate transformations, which is how they will be called in the remainder of the paper. The physics is supposed to be unaffected by those, but merely its description changes. This property becomes manifest in the covariance of GR. A general coordinate transformation does not affect individual points $p\in\mathcal{M}$, but changes coordinate patches, which cover the manifold in a smooth manner. The way how the tensor changes is via suitable contractions of Jacobians of the transformation, e.g.,
\begin{equation}
T'^{\varrho\sigma\dots}(x')=\frac{\partial x'^{\varrho}}{\partial x^{\mu}}\frac{\partial x'^{\sigma}}{\partial x^{\nu}}\dots T^{\mu\nu\dots}(x')\,.
\end{equation}
By writing the coordinate transformation in the form $x'^{\mu}=x^{\mu}-\xi^{\mu}$, a generic tensor is affected in the same way as it is by an active diffeomorphism, i.e., by the Lie derivative $\mathcal{L}_{\xi}T^{\mu\nu\dots}$. The only difference is in the global sign of the vector field \cite{Bluhm:2014oua}.
Finally, in the context of spacetime symmetry violation, there is a third class of nonlinear transformations on the spacetime manifold that must be introduced. They are the particle diffeomorphisms~\cite{Bluhm:2014oua} already mentioned previously.

Now, in the context of spacetime symmetry violations, we must distinguish between two types of tensors. The first type transforms under active diffeomorphisms and general coordinate transformations as expected for a tensor-valued field. Active diffeomorphisms are equivalent to particle diffeomorphisms for these tensors. The second type involves tensor-valued background fields that transform as usual under general coordinate transformations and active diffeomorphisms. However, they do not transform at all under particle diffeomorphisms. In other words, tensor-valued vacuum values transform trivially under these diffeomorphisms. When a particle diffeomorphism acts on $p\in\mathcal{M}$, such a tensor simply does not take any notice of what occurs in the manifold proper. The term \textit{background} describes this behavior accurately.

Note that $u(x)$ is a tensor of rank~0 of the first type, i.e., it transforms under active and particle diffeomorphisms as well as general coordinate transformations in the way expected: $u(x)\mapsto u'(x)=u(x)+\mathcal{L}_{\xi}u(x)$ and $u'(x')=u(x)-\mathcal{L}_{\xi}u(x)$, respectively. On the contrary, the vacuum value $\langle u\rangle$ is a tensor of rank~0 of the second type. Thus, it transforms under active diffeomorphisms in the usual manner, $\langle u\rangle (x)\mapsto \langle u\rangle '(x)=\langle u\rangle(x)+\mathcal{L}_{\xi}\langle u\rangle(x)$. Also, under general coordinate transformations it holds that $\langle u\rangle'(x')=\langle u\rangle (x)-\mathcal{L}_{\xi}\langle u\rangle(x)$, in analogy to $u(x)$. However, the vacuum value exhibits a nonstandard behavior under particle diffeomorphisms: $\langle u\rangle (x)\mapsto \langle u\rangle'(x)=\langle u\rangle(x)$ \cite{Bluhm:2004ep,Bailey:2006fd}.

Without any background field present, the effect of each active/particle diffeomorphism can be reversed by an appropriate general coordinate transformation. Then, the physics before and after this sequence of transformations, which is basically the identity, is the same. The situation changes drastically when there is a background field, as now the effect of a particle diffeomorphism cannot simply be eliminated by a general coordinate transformation. While the sequence of an active diffeomorphism and an appropriately chosen general coordinate transformation provides the identity, this is not the case when the active diffeomorphism is replaced by a particle diffeomorphism. Therefore, the physics before and after the set of transformations may be very different. The particle diffeomorphism generators that are spontaneously broken in the presence of a vacuum value $\langle u\rangle(x)$ are those along vector fields $\xi$ that satisfy $\mathcal{L}_{\xi}\langle u\rangle=\xi^{\mu}\partial_{\mu}\langle u\rangle\neq 0$. These diffeomorphisms cannot be reversed by any general coordinate transformation. In the following, if not stated otherwise, we will be referring to \textit{particle} diffeomorphisms as diffeomorphisms, for brevity.

Speaking of the $(3+1)$ decomposition, one also distinguishes between spacetime diffeomorphisms and spatial diffeomorphisms \cite{Gambini:1998it}. On the one hand, spacetime diffeomorphisms are generated by the particular vector field $m^{\mu}=Nn^{\mu}$, which is introduced via the foliation of $\mathcal{M}$ in the $(3+1)$ decomposition. A generic spacetime tensor $T^{\mu\nu\dots}$, which is \emph{not} a background, changes by $\mathcal{L}_mT^{\mu\nu\dots}$. When it comes to the constraints, it is the Hamiltonian constraint that is associated with spacetime diffeomorphisms, i.e., there is a single generator in this case. On the other hand, spatial diffeomorphisms are generated by the three components of the shift vector. Thus, when a spatial diffeomorphism is performed, $T^{\mu\nu\dots}$ is supposed to change by $\mathcal{L}_{\mathbf{N}}T^{\mu\nu\dots}$. Note that the three momentum constraints are associated with spatial diffeomorphisms such that there are three generators. In total, there are four generators for diffeomorphisms.

These generators are preserved or broken depending on the form of the vacuum value $\langle u\rangle$ as a function of the spacetime coordinates. Table~\ref{tab:generators-diffeomorphisms} provides a summary of characteristic configurations that can occur. Each scenario imaginable for the number of broken generators can be realized. All generators can be preserved, but presumably only for a spacetime-independent $\langle u\rangle$, which is the simplest case possible. However, such a $\langle u\rangle$ could only arise from a constant background field $u$, which is known to not contain physical information~\cite{Bailey:2006fd}. After all, a latter can be absorbed into the gravitational field by a redefinition of the metric.
The more complicated the spacetime dependence of $\langle u\rangle$ gets, the more generators for diffeomorphisms are spontaneously broken. As long as $\langle u\rangle$ only depends on time or certain components of the spatial coordinates, a subset of the generators can be preserved. All generators are broken when $\langle u\rangle$ depends on all the spacetime coordinates.

\begin{table}
\centering
\begin{tabular}{cccccc}
\toprule
$\langle u\rangle$ & $(m^{\mu})=(1,-\mathbf{N})$ & $\#_1$ & $\tilde{\mathbf{N}}$ & $\#_2$ & $\#_1+\#_2$ \\
\midrule
constant & $(1,-N^1,-N^2,-N^3)$ & 1 & $(\tilde{N}^1,\tilde{N}^2,\tilde{N}^3)$ & 3 & 4 \\
$\langle u\rangle(x^1)$ & $(1,0,-N^2,-N^3)$ & 1 & $(0,\tilde{N}^2,\tilde{N}^3)$ & 2 & 3 \\
                        & $(1,-N^1,0,0)$ & 0 & $(\tilde{N}^1,0,0)$ & 0 & 0 \\
$\langle u\rangle(x^1,x^2)$ & $(1,0,0,-N^3)$ & 1 & $(0,0,\tilde{N}^3)$ & 1 & 2 \\
                            & $(1,-N^1,-N^2,0)$ & 0 & $(\tilde{N}^1,\tilde{N}^2,0)$ & 0 & 0 \\
$\langle u\rangle(\mathbf{x})$ & $(1,0,0,0)$ & 1 & $(\tilde{N}^1,\tilde{N}^2,\tilde{N}^3)$ & 0 & 1 \\
                               & $(1,-N^1,-N^2,-N^3)$ & 0 & $(\tilde{N}^1,\tilde{N}^2,\tilde{N}^3)$ & 0 & 0 \\
$\langle u\rangle(t)$ & $(1,-N^1,-N^2,-N^3)$ & 0 & $(\tilde{N}^1,\tilde{N}^2,\tilde{N}^3)$ & 3 & 3 \\
$\langle u\rangle(t,x^1)$ & $(1,-N^1,-N^2,-N^3)$ & 0 & $(0,\tilde{N}^2,\tilde{N}^3)$          & 2 & 2 \\
                          &                      &   & $(\tilde{N}^1,0,0)$ & 0 & 0\\
$\langle u\rangle(t,x^1,x^2)$ & $(1,-N^1,-N^2,-N^3)$ & 0 & $(0,0,\tilde{N}^3)$          & 1 & 1 \\
                          &                          &   & $(\tilde{N}^1,\tilde{N}^2,0)$ & 0 & 0\\
$\langle u\rangle(t,\mathbf{x})$ & $(1,-N^1,-N^2,-N^3)$ & 0 & $(\tilde{N}^1,\tilde{N}^2,\tilde{N}^3)$ & 0 & 0 \\
\bottomrule
\end{tabular}
\caption{Identification of preserved generators for spacetime and spatial diffeomorphisms \cite{Gambini:1998it} via the behaviors of $\mathcal{L}_m\langle u\rangle$ and $\mathcal{L}_{\mathbf{N}}\langle u\rangle$, respectively, for different spacetime dependencies of $\langle u\rangle$. The first column presents the specific configuration for the vacuum value of $\langle u\rangle$ under consideration. The second (fourth) columns contain the vector fields describing spacetime (spatial) diffeomorphisms. The third (fifth) columns state the numbers of preserved generators for spacetime (spatial) diffeomorphisms. Finally, the last column gives the total number of preserved generators. We distinguish between $\mathbf{N}$ in the spatial part of $m^{\mu}$ and the shift vector $\tilde{\mathbf{N}}$, since both types of diffeomorphisms can be studied independently of each other.}
\label{tab:generators-diffeomorphisms}
\end{table}

As we argued before, the presence of a nonzero vacuum value $\langle u\rangle$ also gives rise to a nonzero derivative $\partial_{\mu}\langle u\rangle$, which defines a preferred vector field in a spacetime tangent frame. Via the vierbein, another preferred vector field $V_a:=e_a^{\phantom{a}\mu}\partial_{\mu}\langle u\rangle$ can be inferred to exist in each freely-falling frame. Depending on the form of the latter, the generators of the local Lorentz group are broken accordingly. The form of $V_a$ is not only related to the spacetime dependence of $\langle u\rangle$, but the form of the vierbein $e_a^{\phantom{a}\mu}$ also plays a role. Since a particular form of the vierbein shall not be assumed, we will be referring to the form of $V_a$ only.

In analogy to diffeomorphisms and general coordinate transformations, we must distinguish between local particle and observer Lorentz transformations, respectively, where the latter are, in fact, mere coordinate transformations \cite{Colladay:1998fq}. The components $V_a$ are part of a background field, i.e., they transform as $V_a\mapsto V_a'=V_a$ and $V_a\mapsto V_a'=\Lambda_a^{\phantom{a}b}V_b$ under local particle and observer transformations, respectively. To deduce which local-Lorentz generators are broken, we must investigate the behavior of $\Lambda_a^{\phantom{a}b}V_b$.
\begin{table}
\centering
\begin{tabular}{cccccc}
\toprule
$V_a$ & Boosts $\boldsymbol{\beta}$ & $\#_1$ & Rotations $\boldsymbol{\theta}$ & $\#_2$ & $\#_1+\#_2$ \\
\midrule
$(V_0,0,0,0)$ & $(\beta^j)_j$ & 0 & $(\theta^j)_j$ & 3 & 3 \\
$(0,V_i\delta_{ij})_j$ & $(\beta^j-\beta^i\delta_{ij})_j$ & 2 & $(\theta^i\delta_{ij})_j$ & 1 & 3 \\
$(0,V_j-V_i\delta_{ij})_j$ & $(\beta^i\delta_{ij})_j$ & 1 & $(\theta^j)_j$ & 0 & 1 \\
$(V_0,V_i\delta_{ij})_j$ & $(\beta^j)_j$ & 0 & $(\theta^i\delta_{ij})_j$ & 1 & 1 \\
$(V_0,\mathbf{V})$ & $(\beta^j)_j$ & 0 & $(\theta^j)_j$ & 0 & 0 \\
\bottomrule
\end{tabular}
\caption{Impact of local observer Lorentz transformations on $V_a$. The first column shows the specific form of $V_a$ of interest. The second (fourth) columns state the boost (rotation) parameters. Here, $\beta^i$ indicates the boost velocity along the $i$-th axis where $\theta^i$ stands for the rotation angle around the $i$-th axis. Note that $i\in\{1,2,3\}$ fixed, whereas $j$ runs over $1\dots 3$. The third (fifth) columns provide the numbers of preserved generators for boosts (rotations). Last but not least, the sixth column lists the total number of preserved generators.}
\label{tab:generators-local-lorentz-transformations}
\end{table}

At least 3 out of 6 local-Lorentz generators are broken. For example, an isotropic configuration for $V_a$ leads to all boost generators being broken, whereas each of the rotations is preserved due to the remaining $\mathit{SO}(3)$ invariance in a local frame. Configurations can be chosen that partially preserve the boosts at the cost of breaking some of the generators for rotations. The latter is the case when, e.g., there is a residual local $\mathit{SO}(2)$ symmetry. However, this residual invariance can also be present in cases where boost invariance is lost completely, whereupon 5 Lorentz generators are spontaneously broken. Finally, for generic choices of $V_a$, not a single Lorentz generator is preserved; see table~\ref{tab:generators-local-lorentz-transformations} for a summary of some important configurations.

\section{Field equations in covariant formulation}
\label{sec:field-equations}

Having discussed the existence of a generic vacuum expectation value, we would like to dedicate ourselves to the field equations of eq.~\eqref{eq:action-BD-type-theory} for $\phi=1-u$, i.e., in the covariant formulation before ADM-decomposing the action. The gravitational field equations are obtained by varying the action for the metric, where computational details are shown in appendix~\ref{sec:field-equations-gravitational-field}. The field equations before spontaneous symmetry breaking can be expressed in terms of the Einstein tensor ${}^{(4)}G_{\mu\nu}:={}^{(4)}R_{\mu\nu}-({}^{(4)}R/2)g_{\mu\nu}$ as follows,
\begin{subequations}
\label{eq:modified-einstein-equations-both}
\begin{align}
\label{eq:modified-einstein-equations}
0&=(1-u){}^{(4)}G_{\mu\nu}+\nabla_\mu \nabla_\nu u-g_{\mu\nu}\square u \notag \\
&\phantom{{}={}}-\frac{\omega}{1-u}\bigg(\nabla_\mu u \nabla_\nu u-\frac{1}{2}g_{\mu\nu}\nabla_{\varrho} u \nabla^{\varrho} u\bigg)+\frac{1}{2}g_{\mu\nu}V(u)\,,
\end{align}
and in their trace-reversed form
\begin{align}
\label{eq:modified-einstein-equations-trace-reversed}
0&=(1-u){}^{(4)}R_{\mu\nu}+\nabla_\mu \nabla_\nu u+\frac{1}{2}g_{\mu\nu}\square u \notag \\
&\phantom{{}={}}-\frac{\omega}{1-u}\bigg[\nabla_\mu u \nabla_\nu u+\frac{1}{2}g_{\mu\nu}(\square u-\nabla_{\varrho} u \nabla^{\varrho} u)\bigg]-\frac{1}{2}g_{\mu\nu}V(u)\,,
\end{align}
\end{subequations}
respectively, where $\square:=\nabla^2=\nabla_{\varrho}\nabla^{\varrho}$ is the d'Alembertian.
The latter are interpreted as modified Einstein equations for a gravitational theory in the absence of matter, but supplemented by an additional dynamical scalar field~$u$. The first line of eq.~\eqref{eq:modified-einstein-equations} matches the modified field equations for $u$ stated in ref.~\cite{Bailey:2006fd}. Furthermore, eqs.~\eqref{eq:modified-einstein-equations} and \eqref{eq:modified-einstein-equations-trace-reversed} clearly reduce to the Einstein equations in vacuum for $u=0$ and when the potential $V(u)$ is discarded. The fourth and fifth terms of eq.~\eqref{eq:modified-einstein-equations} partially describe the dynamics of the scalar field. The latter terms are clearly absent in a scenario of explicit spacetime symmetry violation. Note that $u$ transforms in the same way as does any scalar field under general coordinate transformations and diffeomorphisms.

The field equation for the scalar field $u$ is also computed in appendix~\ref{sec:field-equations-u}:
\begin{equation}
\label{eq:field-equation-u}
0={}^{(4)}R-\frac{\omega}{1-u}\left(\frac{1}{1-u}\nabla_{\mu}u\nabla^{\mu}u+2\square u\right)+V'(u)\,.
\end{equation}
Spacetime geometry contributes to the dynamics of $u$ via the first and third terms. The second is purely kinematical and the fourth involves the first derivative of the potential. Interestingly, when $u=0$ and $V'(u)$ is discarded, the field equation reduces to ${}^{(4)}R=0$, which is the trace of the usual Einstein equations in vacuum. Thus, the absence of $u$ then simply leads to a superfluous equation, which does not contain any additional physical information and is not in contradiction with the Einstein equations.

Now, eqs.~\eqref{eq:modified-einstein-equations} and \eqref{eq:field-equation-u} form a system of 11 independent coupled nonlinear partial differential equations, which is obviously challenging to solve analytically without specifying a particular gravitational system. We can still make some generic statements, though. As long as $u$ has not reached the global minimum of the potential $V(u)$, we can interpret $V'(u)$ as a nonvanishing force driving the scalar background field towards its vacuum expectation value. The kinematic contributions for $u$ as well as the gravitational field are expected to change in this process. After reaching the vacuum value $\langle u\rangle$, we have $V'(\langle u\rangle)=0$, whereupon eqs.~\eqref{eq:modified-einstein-equations} and \eqref{eq:field-equation-u} take the form
\begin{subequations}
\begin{align}
\label{eq:modified-einstein-equation-after-SSB}
0&=(1-\langle u\rangle){}^{(4)}G_{\mu\nu}+\nabla_\mu \nabla_\nu \langle u\rangle-g_{\mu\nu}\square\langle u\rangle \notag \\
&\phantom{{}={}}-\frac{\omega}{1-u}\bigg(\nabla_\mu \langle u\rangle \nabla_\nu \langle u\rangle-\frac{1}{2}g_{\mu\nu}\nabla_{\varrho} \langle u\rangle \nabla^{\varrho} \langle u\rangle\bigg)+\frac{1}{2}g_{\mu\nu}V(\langle u\rangle)\,, \\[1ex]
0&={}^{(4)}R-\frac{\omega}{1-\langle u\rangle}\left(\frac{1}{1-\langle u\rangle}\nabla_{\mu}\langle u\rangle\nabla^{\mu}\langle u\rangle+2\square\langle u\rangle\right)\,.
\end{align}
\end{subequations}
After $u$ has picked a specific $\langle u\rangle$ at each spacetime point dynamically, it corresponds to a fixed background field permeating the vacuum with the gravitational field present. When this occurs, the presence of $\langle u\rangle$ in the first three terms of eq.~\eqref{eq:modified-einstein-equations} indicates that any solution to these equations will be modified as compared to GR. A solution is given by a metric $g_{\mu\nu}=g_{\mu\nu}(x^{\varrho},\langle u\rangle)$, which clearly depends on the vacuum value of $u$. Thus, any gravitational field described by the connection coefficients $\Gamma^{\mu}_{\phantom{\mu}\nu\varrho}$ is expected to be distorted by $\langle u\rangle$. The kinetic and potential terms, which are now evaluated for $\langle u\rangle$, will contribute to modifying the gravitational field. In principle, the potential can be shifted along the energy axis such that $V(\langle u\rangle)=0$. Then, the last term of eq.~\eqref{eq:modified-einstein-equation-after-SSB} is eliminated.

Note also that eqs.~\eqref{eq:modified-einstein-equations-both} and \eqref{eq:field-equation-u} are perfectly consistent with the contracted second Bianchi identities $\nabla^{\mu}{}^{(4)}G_{\mu\nu}=0$. For example, computing the divergence of eq.~\eqref{eq:modified-einstein-equations}, we have
\begin{align}
\label{eq:check-consistency-dynamics}
0&=\frac{1}{2}{}^{(4)}R\nabla_{\nu}u-\frac{\omega}{(1-u)^2}\left(\nabla^{\mu}u\nabla_{\mu}u\nabla_{\nu}u-\frac{1}{2}\nabla_{\nu}u\nabla_{\varrho}u\nabla^{\varrho}u\right) \notag \\
&\phantom{{}={}}-\frac{\omega}{1-u}\left(\square u\nabla_{\nu}u+\nabla_{\mu}u\nabla^{\mu}\nabla_{\nu}u-\nabla_{\nu}\nabla_{\varrho}u\nabla^{\varrho}u\right)+\frac{1}{2}V'(u)\nabla_{\nu}u \notag \\
&=\frac{1}{2}{}^{(4)}R\nabla_{\nu}u-\frac{\omega}{2(1-u)^2}\nabla_{\mu}u\nabla^{\mu}u\nabla_{\nu}u-\frac{\omega}{1-u}\square u\nabla_{\nu}u+\frac{1}{2}V'(u)\nabla_{\nu}u\,,
\end{align}
by using the first of eq.~(6) in ref.~\cite{Bailey:2024zgr} and $[\nabla_{\mu},\nabla_{\nu}]u=0$. This result corresponds to eq.~\eqref{eq:field-equation-u} when multiplied with $(1/2)\nabla_{\nu}u$. Thus, pseudo-Riemannian geometry does not clash with the dynamics. However, in the case of explicit spacetime symmetry breaking, the kinetic and potential terms of eq.~\eqref{eq:modified-einstein-equations-both} are absent, so is eq.~\eqref{eq:field-equation-u}. The situation is then more involved and requires alternative methods for a consistent treatment; see, in particular, the recent ref.~\cite{Reyes:2024ywe}.

\subsection{Physical interpretation of diffeomorphism violation and phenomenology}

Imagine a gravitational field with $\langle u\rangle=0$, where a number of laboratories are placed at fixed positions such that experimentalists perform mechanical, electromagnetic, and optical experiments locally. According to Einstein's equivalence principle, these laboratories are equivalent to accelerated frames without gravity present. An active/particle diffeomorphism is interpreted to translate laboratories to different positions within the gravitational field, where the underlying spacetime manifold is mapped onto itself. Measurements performed in the rearranged laboratories then still provide results consistent with the gravitational field present, since a diffeomorphism should not affect the physics of GR, after all.

Now, when $\langle u\rangle$ is switched on, it must be considered as a background function on the spacetime manifold. A particle diffeomorphism acts on the experiments as does an active diffeomorphism, but $\langle u\rangle$ remains unaffected. Translating an experiment from a point $x_1$ to $x_2$, the vacuum value $\langle u\rangle(x_1)$ does not move together with the experiment, but adapts to its value $\langle u\rangle(x_2)$. Thus, when the experimentalist tries to relate their new results measured at $x_2$ to their previous results at $x_1$, they will find disagreements with what would be expected from GR. This is how spontaneous diffeomorphism violation becomes manifest phenomenologically. The spacetime metric $g_{\mu\nu}(x^{\varrho},\langle u\rangle)$ is pseudo-Riemannian, since the symmetry-violating background arises dynamically (see the discussion around the previous eq.~\eqref{eq:check-consistency-dynamics}). In contrast to explicit symmetry violation, neither does a beyond-Riemannian setting arise \cite{Kostelecky:2020hbb,Kostelecky:2021tdf} nor is it necessary to reduce spacetime geometry such as described in ref.~\cite{Reyes:2024ywe}.

Now, scalar-tensor gravity theories have already been severely constrained by experiments, e.g., based on the Cassini spacecraft \cite{Bertotti:2003rm} and binary star systems \cite{Freire:2012mg}. However, these constraints have to be interpreted in the local context they were obtained in. Experimental tests of GR as they are described in the latter papers are well suited for testing GR at comparably small length scales, where $\langle u\rangle$ could basically be independent of the spacetime coordinates. Note that a constant $\langle u\rangle$ does not give rise to spacetime symmetry violation (see table~\ref{tab:generators-diffeomorphisms}), which renders such a setting trivial. On the contrary, experiments of the kind of refs.~\cite{Bertotti:2003rm,Freire:2012mg} are not expected to rule out a vacuum value $\langle u\rangle$ that changes significantly at cosmological scales only.

\section{Hamiltonian formulation}
\label{sec:hamiltonian-formulations}

Our interest in the remainder of the paper is to set up the Hamiltonian formulation of eq.~\eqref{eq:u-action-ADM-decomposed} in the first thermodynamic phase, i.e., before spontaneous symmetry breaking. When the background field does not arise dynamically, i.e., spacetime symmetry breaking is explicit, the constraint structure of the resulting modified-gravity theory is known to be modified substantially \cite{Reyes:2021cpx,Reyes:2022mvm,Reyes:2023sgk}. Therefore, the number of physical (propagating) degrees of freedom can differ from that of GR \cite{Kostelecky:2017zob,ONeal-Ault:2020ebv,Bailey:2024zgr}. Our purpose in the following is to determine all the constraints, the full Hamiltonian of the theory, and the number of physical (propagating) degrees of freedom for the case of the dynamical scalar field $u$, which gives rise to a background dynamically.

To do so, we need the canonical momentum densities associated with the dynamical variables. In the case of explicit spacetime symmetry violation as studied in refs.~\cite{Reyes:2021cpx,Reyes:2022mvm,Reyes:2023sgk}, the dynamical configuration space variables are the metric components $q_{ab}$ on $\Sigma_t$ with
\begin{equation}
\pi^{ab}:=\frac{\partial\mathcal{L}^{(3+1)}}{\partial\dot{q}_{ab}}\,,
\end{equation}
being the components of the canonical momentum density associated with $q_{ab}$. In particular, for eq.~\eqref{eq:u-action-ADM-decomposed} we have that
\begin{subequations}
\begin{equation}
\pi^{ab}=\frac{\sqrt{q}}{2\kappa}\left[(1-u)(K^{ab}-Kq^{ab})+\frac{q^{ab}}{N}(\dot{u}-N^cD_cu)\right]\,,
\end{equation}
as well as for the trace $\pi:=q_{ab}\pi^{ab}$:
\begin{equation}
\label{eq:canonical-momentum-trace}
\pi=\frac{\sqrt{q}}{2\kappa}\left[-2(1-u)K+\frac{3}{N}(\dot{u}-N^cD_cu)\right]\,.
\end{equation}
\end{subequations}
A crucial difference to our approach in refs.~\cite{Reyes:2021cpx,Reyes:2022mvm,Reyes:2023sgk} is that there are additional dynamical degrees of freedom such as $u$. Therefore, these come with canonical momenta, which we must compute. Now, based on eqs.~\eqref{eq:lagrange-density-R-u}, \eqref{eq:lagrange-density-kinematics-u}, we introduce the canonical momentum associated with $u$ as follows:
\begin{equation}
\label{eq:canonical-momentum-u}
\pi_u:=\frac{\partial\mathcal{L}^{(3+1)}}{\partial\dot{u}}=\frac{\sqrt{q}}{2\kappa}\left[\frac{2\omega}{N(1-u)}(\dot{u}-N^aD_au)+2K\right]\,.
\end{equation}
So the latter is directly related to the kinetic term of the Lagrange density, which is expected. Furthermore, the way how $\Sigma_t$ is embedded into $\mathcal{M}$ also contributes to the canonical momentum via the last term of $\pi_u$. Thus, from eqs.~\eqref{eq:canonical-momentum-trace}, \eqref{eq:canonical-momentum-u} we observe that $\pi_u=-\pi$ for $u=0$. So even when the scalar background is discarded, the associated canonical momentum $\pi_u$ does not vanish, which will be important in the following. Recall from section~\ref{sec:field-equations} that the covariant field equation for $u$ reduces to ${}^{(4)}R=0$ for $u=0$, which is a superfluous, but not a trivial equation.

To arrive at the Hamiltonian associated with eq.~\eqref{eq:u-action-ADM-decomposed}, the configuration space variables must be expressed in terms of the canonical momenta associated. First of all, the trace of the extrinsic curvature as well at the extrinsic curvature proper are written in terms of $\pi^{ab}$ and its trace as follows:
\begin{subequations}
\begin{align}
\label{eq:extrinsic-curvature-scalar-canonical-momentum}
K&=\frac{1}{2(1-u)}\left[\frac{3}{N}(\dot{u}-N^aD_au)-\frac{2\kappa}{\sqrt{q}}\pi\right]\,, \\[2ex]
\label{eq:extrinsic-curvature-tensor-canonical-momentum}
K^{ab}&=\frac{1}{1-u}\left[\frac{2\kappa}{\sqrt{q}}\left(\pi^{ab}-\frac{\pi}{2}q^{ab}\right)+\frac{q^{ab}}{2N}(\dot{u}-N^cD_cu)\right]\,.
\end{align}
\end{subequations}
Note that the canonical momentum of $u$ in eq.~\eqref{eq:canonical-momentum-u} still involves the extrinsic-curvature scalar, which should be eliminated via eq.~\eqref{eq:extrinsic-curvature-scalar-canonical-momentum}:
\begin{equation}
\pi_u=\frac{1}{1-u}\left[\frac{\sqrt{q}}{2\kappa N}(3+2\omega)(\dot{u}-N^aD_au)-\pi\right]\,.
\end{equation}
The latter can now be inverted for $\dot{u}$:
\begin{equation}
\label{eq:time-derivative-u-canonical-momentum}
\dot{u}=\frac{2\kappa}{\sqrt{q}}\frac{N}{3+2\omega}\big[\pi+(1-u)\pi_u\big]+N^aD_au\,.
\end{equation}
Hence, the extrinsic curvature of eq.~\eqref{eq:extrinsic-curvature-tensor-canonical-momentum} expressed completely in terms of canonical momenta is
\begin{equation}
\label{eq:extrinsic-curvature-tensor-complete}
K^{ab}=\frac{\kappa}{\sqrt{q}(1-u)}\bigg\{2\left(\pi^{ab}-\frac{\pi}{2}q^{ab}\right)+\frac{q^{ab}}{3+2\omega}\big[\pi+(1-u)\pi_u\big]\bigg\}\,.
\end{equation}
The Hamilton density arises from the Legendre transformation of the Lagrange density, which must be performed with respect to $q_{ab}$ and $u$:
\begin{equation}
\mathcal{H}_u=\pi^{ab}\dot{q}_{ab}+\pi_u\dot{u}-\mathcal{L}^{(3+1)}_u\,,
\end{equation}
i.e., we must also express $\dot{q}_{ab}$ via $\pi^{ab}$ and $\pi_u$. To do so, we employ eq.~\eqref{eq:extrinsic-curvature-tensor-complete} as well as the defining relationship of the extrinsic-curvature tensor,
\begin{equation}
\label{eq:defining-relation-extrinsic-curvature}
\dot{q}_{ab}=2NK_{ab}+D_aN_b+D_bN_a\,.
\end{equation}
The latter remains standard, since, after all, eq.~\eqref{eq:u-action-ADM-decomposed} rests upon pseudo-Riemannian geometry, as we discussed before in detail.
Finally, the Hamiltonian formulation of eq.~\eqref{eq:u-action-ADM-decomposed} is governed by
\begin{subequations}
\label{eq:hamiltonian-scalar-model}
\begin{align}
H_u&=\int_{\Sigma_t} \mathrm{d}^3y\,\mathcal{H}_u\,, \\[2ex]
\mathcal{H}_u&=\frac{\sqrt{q}}{2\kappa}\bigg\{N\bigg[-(1-u)R+\frac{4\kappa^2}{(1-u)q}\left(\pi_{ab}\pi^{ab}-\frac{\pi^2}{2}\right)+\frac{2\kappa^2}{(1-u)q}\frac{1}{3+2\omega}\big[\pi+(1-u)\pi_u\big]^2 \notag \\
&\phantom{{}={}}\hspace{1.3cm}+\frac{\omega}{1-u}D_auD^au+V(u)\bigg]+2(1-u)D^aD_aN\bigg\}+2\pi_a^{\phantom{a}b}D_bN^a+\pi_uN^aD_au\,.
\end{align}
\end{subequations}
In the recent paper \cite{Reyes:2023sgk} we described how to deal with integrations by parts and boundary terms properly. Although the latter analysis was performed for a theory with explicit spacetime symmetry violation, the particular aspects related to boundary terms is expected to be independent of whether symmetries are violated explicitly or spontaneously. After all, the kinetic and potential terms of $u$ are not integrated by parts, but only some of the remaining terms are. Integrations by parts are carried out to get rid of derivatives acting on the lapse function and the shift vector, respectively, such that the Hamiltonian is recast into the alternative form
\begin{subequations}
\label{eq:hamiltonian-scalar-model-after-integrations-by-parts}
\begin{align}
\tilde{H}_u&=\int_{\Sigma_t} \mathrm{d}^3y\,(N\mathcal{H}+N^c\mathcal{H}_c)\,, \\[2ex]
\label{eq:hamiltonian-constraint}
\mathcal{H}&=\frac{\sqrt{q}}{2\kappa}\bigg\{-(1-u)R+\frac{4\kappa^2}{(1-u)q}\left(\pi_{ab}\pi^{ab}-\frac{\pi^2}{2}\right)+\frac{2\kappa^2}{(1-u)q}\frac{1}{3+2\omega}\big[\pi+(1-u)\pi_u\big]^2 \notag \\
&\phantom{{}={}}\hspace{0.8cm}+\frac{\omega}{1-u}D_auD^au+V(u)-2D^aD_au\bigg\}\,, \\[2ex]
\label{eq:momentum-constraint}
\mathcal{H}_a&=\pi_uD_au-2D_b\pi_a^{\phantom{a}b}\,.
\end{align}
\end{subequations}
Several comments are in order. First, $\tilde{H}_u$ reproduces the ADM Hamiltonian of GR for vanishing symmetry violation, which means that $u=0$, $\omega=0$, and $\pi_u=-\pi$ (see comment under eq.~\eqref{eq:canonical-momentum-u}). Second, $\tilde{H}_u$ preserves the overall structure of the ADM Hamiltonian for GR, i.e, the modified Hamiltonian decomposes into a contribution multiplied by the lapse function and another one contracted with the shift vector components. This property is crucial and an implication of $u$ being a dynamical field. After all, eq.~\eqref{eq:u-action-ADM-decomposed} is still invariant under spacetime and spatial diffeomorphisms. However, they can be broken by the ground state of $u$, i.e., the vacuum expectation value $\langle u\rangle$ that the background field chooses in the potential $V(u)$; see section~\ref{sec:potential}.

So eq.~\eqref{eq:hamiltonian-constraint} is concluded to only involve mild modifications of GR. Note that when diffeomorphism symmetry is broken explicitly, the overall structure of the modified Hamiltonian exhibits crucial differences from that of GR. In this case new terms emerge that are neither linear in the lapse function nor the shift vector \cite{Reyes:2021cpx,Reyes:2022mvm,Reyes:2023sgk}. Such changes are severe and imply significant deviations from GR, in particular, when it comes to its constraint structure. The forthcoming two sections will be dedicated to a discussion of these aspects.

\section{Symmetry generators, field equations, and constraints}
\label{sec:symmetry-generators-field-equations}

Let $X\in \{N,N^a,q_{ab},u\}$ be the canonical variables and $\Pi\in \{\pi_N,\pi_a,\pi^{ab},\pi_u\}$ the corresponding canonical momenta. The symmetries present in eq.~\eqref{eq:action-BD-type-theory} are encoded in Poisson brackets between suitably chosen quantities. In general, let us consider two quantities $Q(\mathbf{x})$ and $P(\mathbf{x})$ given on $\Sigma_t$. The Poisson bracket of the latter is then defined as
\begin{align}
\{Q(\tilde{x},\tilde{t}),P(x',t')\}&:=\int_{\Sigma_t}\mathrm{d}^3y\,\left[\frac{\delta Q(\tilde{x},\tilde{t})}{\delta X_c(y,t)}\frac{\delta P(x',t')}{\delta\Pi^c(y,t)}-\frac{\delta Q(\tilde{x},\tilde{t})}{\delta\Pi^c(y,t)}\frac{\delta P(x',t')}{\delta X_c(y,t)}\right]\,,
\end{align}
with the functional derivatives $\delta/\delta X_c$, $\delta/\delta \Pi^c$ for the canonical variables and momenta, respectively. Now, eq.~\eqref{eq:u-action-ADM-decomposed} neither involves $\dot{N}$ nor $\dot{N}^a$. Thus,
\begin{subequations}
\label{eq:canonical-momenta-lapse-shift}
\begin{align}
\pi_N:=\frac{\partial\mathcal{L}_u^{(3+1)}}{\partial\dot{N}}\approx 0\,, \\[2ex]
\pi_a:=\frac{\partial\mathcal{L}_u^{(3+1)}}{\partial\dot{N}^a}\approx 0\,,
\end{align}
\end{subequations}
where `$\approx$' means ``weakly equal to zero.'' Equation~\eqref{eq:canonical-momenta-lapse-shift} must be considered as primary constraints. Obviously, these constraints emerge as long as the ADM-decomposed action does not depend on time derivatives of the lapse function and the shift vector. This is the case in GR as well as in eq.~\eqref{eq:BD-action-ADM-decomposed} where spacetime symmetries are violated spontaneously. However, it can also hold in modified-gravity theories with explicit symmetry violation, as we showed in ref.~\cite{Reyes:2021cpx}. For $\pi_N$ and $\pi_a$ being constrained in the form of eq.~\eqref{eq:canonical-momenta-lapse-shift} it is critical to add suitable boundary terms to the action such that time derivatives $\dot{N}$ and $\dot{N}^a$ can be gotten rid of via suitable integrations by parts. Thus, the nature of symmetry violation is unessential in this particular context. Note that settings were also considered where eq.~\eqref{eq:canonical-momenta-lapse-shift} does not hold, which immediately leads to drastic changes of the constraint structure as compared to GR \cite{ONeal-Ault:2020ebv}.

Now, the above constraints must be adjoined to the canonical Hamiltonian of eq.~\eqref{eq:hamiltonian-scalar-model} via Lagrange multipliers $\lambda$ and $\lambda^a$, respectively. Doing so implies an augmented Hamiltonian:
\begin{equation}
\label{eq:augmented-hamiltonian}
H_A=\int_{\Sigma_t}\mathrm{d}^3y\,(N\mathcal{H}+N^c\mathcal{H}_c+\lambda\pi_N+\lambda^c\pi_c)\,,
\end{equation}
with $\mathcal{H}$ and $\mathcal{H}_c$ of eq.~\eqref{eq:hamiltonian-constraint} and \eqref{eq:momentum-constraint}, respectively. In general, the constraint structure should be preserved as a function of time such that a constraint remains a constraint. The time evolution of a generic constraint $\phi_m$ is governed by
\begin{equation}
\label{eq:time-evolution-constraints}
\frac{\mathrm{d}\phi_m}{\mathrm{d}t}=\{\phi_m,H_A\}\,,
\end{equation}
i.e., the augmented Hamiltonian generates their time evolution. Requiring that $\phi_m$ be stationary implies a secondary constraint:
\begin{equation}
\{\phi_m,H_A\}\approx 0\,.
\end{equation}
Applying this approach to $\pi_N\approx 0$ and $\pi_a\approx 0$ leads to
\begin{subequations}
\label{eq:secondary-constraints}
\begin{align}
\{\pi_N,H_A\}&=-\mathcal{H}\approx 0\,, \\[2ex]
\{\pi_a,H_A\}&=-\mathcal{H}_a\approx 0\,.
\end{align}
\end{subequations}
Hence, $\mathcal{H}$ and $\mathcal{H}_a$ play roles analogous to the Hamiltonian and momentum constraints in GR. However, note that they are modified in our setting by the presence of the scalar field~$u$.

Generators formally describe any continuous (infinitesimal) symmetry transformation. One can either represent generators in a geometrical setting or characterize the latter by dynamical variables and constraints. For example, on the one hand, purely timelike diffeomorphisms, which correspond to active time translations in Minkowski spacetime, are generated by a purely timelike vector field. On the other hand, eq.~\eqref{eq:time-evolution-constraints} tells us also that the augmented Hamiltonian $H_A$ generates these types of diffeomorphisms via an operation that is the Poisson bracket.

In what follows, we indent to understand more about these symmetry transformations. We define the smeared Hamiltonian and momentum constraints as functionals of the lapse function and the shift vector, respectively:
\begin{subequations}
\label{eq:constraints-smeared}
\begin{align}
    \label{eq:hamiltonian-constraint-smeared}
    H[N]&:=\int_{\Sigma_t} \mathrm{d}^3y\,N(y)\mathcal{H}(y)\,, \\[2ex]
    \label{eq:momentum-constraint-smeared}
    P[\mathbf{N}]&:=\int_{\Sigma_t} \mathrm{d}^3y\,N^c(y)\mathcal{H}_c(y)\,.
\end{align}
\end{subequations}
Let us first look at Poisson brackets of the dynamical variables with the smeared momentum constraint:
\begin{subequations}
\label{eq:poisson-brackets-momentum-constraint}
\allowdisplaybreaks
\begin{align}
\label{eq:poisson-bracket-momentum-constraint-q}
\{q_{ab},P[\mathbf{N}]\}&=q_{bc}D_aN^c+q_{ac}D_bN^c=\mathcal{L}_{\mathbf{N}}q_{ab}\,, \displaybreak[0]\\[2ex]
\label{eq:poisson-bracket-canonical-momentum}
\{\pi^{ab},P[\mathbf{N}]\}&=N^cD_c\pi^{ab}+\pi^{ab}D_cN^c-2D_cN^{(a}\pi^{b)c}=\mathcal{L}_{\mathbf{N}}\pi^{ab}\,, \displaybreak[0]\\[2ex]
\label{eq:poisson-bracket-momentum-constraint-u}
\{u,P[\mathbf{N}]\}&=N^cD_cu=\mathcal{L}_{\mathbf{N}}u\,, \displaybreak[0]\\[2ex]
\label{eq:poisson-bracket-momentum-constraint-piu}
\{\pi_u,P[\mathbf{N}]\}&=\pi_uD_cN^c+N^cD_c\pi_u=\mathcal{L}_{\mathbf{N}}\pi_u\,,
\end{align}
\end{subequations}
where pairs of indices enclosed in parentheses indicate that this expression must be symmetrized in these indices. Moreover, the unsmeared quantities depend on the spacetime coordinate $x$, which is suppressed, for brevity. From these, eq.~\eqref{eq:poisson-bracket-canonical-momentum} is the most challenging one to compute. Here we employed the Palatini identity, symmetrized the expression with respect to $a,b$, and performed integrations by parts. Interestingly, each one of these results can be expressed as a Lie derivative of a canonical variable with respect to the vector field $\mathbf{N}$ where all Lie derivatives are written in terms of the $q_{ab}$-compatible covariant derivative. A property worthwhile to mention is that $\mathcal{L}_{\mathbf{N}}\pi^{ab}$ involves two more terms compared to $\mathcal{L}_{\mathbf{N}}q_{ab}$. The first is straightforwardly explained by taking into account that the metric $q_{ab}$ is covariantly conserved. The second is explained by noticing that $\pi^{ab}$ is not a mere four-tensor of rank 2, but it is a tensor density. Thus, an additional contribution arises for the Lie derivative because of the factor $\sqrt{q}$, which is implicitly contained in $\pi^{ab}$; see, e.g., ref.~\cite{Yano:1957}.

Geometrically, the shift vector generates spatial diffeomorphisms in the purely spacelike hypersurfaces $\Sigma_t$, which are the analogs of spatial translations in Minkowski spacetime. Moreover, we can also say that the integrated modified momentum constraint generates these spatial diffeomorphisms. This property holds in GR and it continues being valid for eq.~\eqref{eq:action-BD-type-theory} with the dynamical scalar field $\phi=1-u$.

Furthermore, let us take a look at Poisson brackets of the same dynamical variables considered previously with the smeared Hamiltonian constraint:
\begin{subequations}
\allowdisplaybreaks
\begin{align}
\label{eq:poisson-bracket-hamiltonian-constraint-q}
\{q_{ab},H[N]\}&=\frac{2\kappa N}{\sqrt{q}(1-u)}\left(2\left(\pi_{ab}-\frac{\pi}{2}q_{ab}\right)+\frac{q_{ab}}{3+2\omega}\big[\pi+(1-u)\pi_u\big]\right) \notag \displaybreak[0]\\
&=2NK_{ab}=\mathcal{L}_mq_{ab}\,, \\[2ex]
\label{eq:poisson-bracket-hamiltonian-constraint-pi}
\{\pi^{ab},H[N]\}&=\frac{\sqrt{q}}{2\kappa}\Big(-(1-u)NG^{ab}+D^aD^b[(1-u)N]-q^{ab}D^cD_c[(1-u)N]\Big) \notag \displaybreak[0]\\
&\phantom{{}={}}-\frac{2\kappa N}{(1-u)\sqrt{q}}\left[2\left(\pi^{ac}\pi^b_{\phantom{b}c}-\frac{\pi}{2}\pi^{ab}\right)-\frac{q^{ab}}{2}\left(\pi_{cd}\pi^{cd}-\frac{\pi^2}{2}\right)\right] \notag \displaybreak[0]\\
&\phantom{{}={}}-\frac{\kappa N}{\sqrt{q}}\frac{1}{3+2\omega}\left(\frac{\pi}{1-u}+\pi_u\right)\left\{2\pi^{ab}-\frac{q^{ab}}{2}\big[\pi+(1-u)\pi_u\big]\right\} \displaybreak[0]\\
&\phantom{{}={}}-\frac{\sqrt{q}}{2\kappa}\bigg[\frac{N\omega}{1-u}\left(\frac{q^{ab}}{2}D_cuD^cu-D^auD^bu\right) \notag \displaybreak[0]\\
&\phantom{{}={}}\hspace{1.2cm}+q^{ab}\left(D^cND_cu+\frac{N}{2}V(u)\right)-(D^aND^bu+D^bND^au)\bigg]\,, \\[2ex]
\label{eq:poisson-bracket-hamiltonian-constraint-u}
\{u,H[N]\}&=\frac{2\kappa N}{\sqrt{q}}\frac{1}{3+2\omega}\big[\pi+(1-u)\pi_u\big]=\mathcal{L}_mu\,, \\[2ex]
\label{eq:poisson-bracket-hamiltonian-constraint-piu}
\{\pi_u,H[N]\}&=-\frac{N\sqrt{q}}{2\kappa}R-\frac{2\kappa N}{(1-u)^2\sqrt{q}}\left(\pi_{ab}\pi^{ab}-\frac{\pi^2}{2}\right)-\frac{\kappa N}{\sqrt{q}}\frac{1}{3+2\omega}\left(\frac{\pi^2}{(1-u)^2}-\pi_u^2\right) \notag \displaybreak[0]\\
&\phantom{{}={}}+\frac{\sqrt{q}}{\kappa}\left(\omega\left[D_a\left(\frac{ND^au}{1-u}\right)-\frac{N}{2}D_a\left(\frac{1}{1-u}\right)D^au\right]+D_aD^aN\right) \notag \displaybreak[0]\\
&\phantom{{}={}}-N(y)V'(u)\,,
\end{align}
\end{subequations}
where we resort to eq.~\eqref{eq:extrinsic-curvature-tensor-complete}. Also, eq.~\eqref{eq:time-derivative-u-canonical-momentum} is used to derive eq.~\eqref{eq:poisson-bracket-hamiltonian-constraint-u}. Equations~\eqref{eq:poisson-bracket-hamiltonian-constraint-q} and \eqref{eq:poisson-bracket-hamiltonian-constraint-u} show that the Hamiltonian constraint generates spacetime diffeomorphisms for $q_{ij}$ and $u$, respectively, where these are described by Lie derivatives along the vector field $m^{\mu}$. By adding eqs.~\eqref{eq:poisson-bracket-momentum-constraint-q} and \eqref{eq:poisson-bracket-hamiltonian-constraint-q} we deduce the first set of Hamilton's equations for~$q_{ab}$:
\begin{equation}
\label{eq:identity-geometrical}
\dot{q}_{ab}=\left\{q_{ab}(x),\int_{\Sigma_t}\mathrm{d}^3y\,\tilde{\mathcal{H}}_u(y)\right\}\,.
\end{equation}
The latter, when put on paper in their explicit form, correspond to eq.~\eqref{eq:defining-relation-extrinsic-curvature}. Hence, the Hamiltonian formulation incorporates this defining relation, which is a reflection of the unmodified pseudo-Riemannian geometrical setting, which eq.~\eqref{eq:u-action-ADM-decomposed} is based on. Similarly, by adding eq.~\eqref{eq:poisson-bracket-momentum-constraint-u} and eq.~\eqref{eq:poisson-bracket-hamiltonian-constraint-u} we arrive at
\begin{equation}
\label{eq:identity-u}
\dot{u}=\left\{u(x),\int_{\Sigma_t}\mathrm{d}^3y\,\tilde{\mathcal{H}}_u(y)\right\}\,,
\end{equation}
which corresponds to eq.~\eqref{eq:time-derivative-u-canonical-momentum}. This equation is an identity satisfied by $u$ and rests upon the intrinsic geometry on $\Sigma_t$.

In contrast, eqs.~\eqref{eq:poisson-bracket-hamiltonian-constraint-pi}, \eqref{eq:poisson-bracket-hamiltonian-constraint-piu} are of different nature and correspond to the dynamical field equations for the intrinsic metric $q_{ab}$ and the dynamical scalar field $u$, as we shall see shortly. By adding eqs.~\eqref{eq:poisson-bracket-canonical-momentum}, \eqref{eq:poisson-bracket-hamiltonian-constraint-pi} as well as eqs.~\eqref{eq:poisson-bracket-momentum-constraint-piu}, \eqref{eq:poisson-bracket-hamiltonian-constraint-piu}, respectively, we end up with
\begin{subequations}
\begin{align}
\label{eq:dynamical-field-equations-pi}
\dot{\pi}^{ab}&=\left\{\pi^{ab}(x),\int_{\Sigma_t}\mathrm{d}^3y\,\tilde{\mathcal{H}}_u(y)\right\}\,, \displaybreak[0]\\[2ex]
\label{eq:dynamical-field-equations-u}
\dot{\pi}_u&=\left\{\pi_u(x),\int_{\Sigma_t}\mathrm{d}^3y\,\tilde{\mathcal{H}}_u(y)\right\}\,.
\end{align}
\end{subequations}
These are the second sets of Hamilton's equations for the canonical momenta $\pi^{ab}$ and $\pi_u$ associated with $q_{ab}$ and $u$, respectively. In fact, unlike eqs.~\eqref{eq:identity-geometrical}, \eqref{eq:identity-u}, they are the dynamical field equations of the theory~\eqref{eq:action-BD-type-theory} for $\phi=1-u$, which physical field configurations must obey. In the GR limit, i.e., $u=0$, eq.~\eqref{eq:dynamical-field-equations-pi} reduces to the dynamical part of the Einstein equations, whereas eq.~\eqref{eq:dynamical-field-equations-u} reads
\begin{equation}
\label{eq:limit-field-equation-u}
\dot{\pi}_u=-\frac{N\sqrt{q}}{2\kappa}R-\frac{2\kappa N}{\sqrt{q}}\left(\pi_{ab}\pi^{ab}-\frac{\pi^2}{2}\right)+\frac{\sqrt{q}}{\kappa}D_aD^aN+\pi_uD_cN^c+N^cD_c\pi_u\,.
\end{equation}
To understand the meaning of the latter, we resort to the $(3+1)$ decomposition of the Ricci scalar expressed in terms of the canonical momentum; see~eq.~\eqref{eq:ricci-scalar-decomposition}:
\begin{align}
{}^{(4)}R&=\frac{2}{N}\mathcal{L}_mK-\frac{2}{N}D_iD^iN+R+K^2+K_{ab}K^{ab} \notag \\
&=\frac{2}{N}\mathcal{L}_m\left(-\frac{\kappa}{\sqrt{q}}\pi\right)-\frac{2}{N}D_cD^cN+R+\frac{\kappa^2}{q}\pi^2+\frac{4\kappa^2}{q}\left(\pi^{ab}\pi_{ab}-\frac{\pi^2}{4}\right) \notag \\
&=-\frac{2\kappa}{N\sqrt{q}}\mathcal{L}_m\pi-\frac{2\kappa}{N}\pi\mathcal{L}_m\left(\frac{1}{\sqrt{q}}\right)-\frac{2}{N}D_cD^cN+R+\frac{4\kappa^2}{q}\pi^{ab}\pi_{ab}\,,
\end{align}
where the second term can be recast via
\begin{equation}
\mathcal{L}_m\left(\frac{1}{\sqrt{q}}\right)=-\frac{1}{2q^{3/2}}\mathcal{L}_mq=\frac{1}{2q^{3/2}}2q\frac{N}{2}\pi=\frac{N}{2\sqrt{q}}\pi\,.
\end{equation}
Solving ${}^{(4)}R=0$ for $\mathcal{L}_m\pi$ amounts to
\begin{align}
\mathcal{L}_m\pi&=\frac{N\sqrt{q}}{2\kappa}R+\frac{2\kappa N}{\sqrt{q}}\pi^{ab}\pi_{ab}-\frac{\sqrt{q}}{\kappa}D_cD^cN-\frac{\kappa}{\sqrt{q}}\pi^2 \notag \\
&=\frac{N\sqrt{q}}{2\kappa}R+\frac{2\kappa N}{\sqrt{q}}\left(\pi^{ab}\pi_{ab}-\frac{\pi^2}{2}\right)-\frac{\sqrt{q}}{\kappa}D_cD^cN\,.
\end{align}
Finally, we rewrite the Lie derivative of $\pi$ as
\begin{equation}
\mathcal{L}_m\pi=\dot{\pi}-\mathcal{L}_{\mathbf{N}}\pi=\dot{\pi}-\pi D_cN^c-N^cD_c\pi\,, \\[1ex]
\end{equation}
to obtain
\begin{equation}
\dot{\pi}=\frac{N\sqrt{q}}{2\kappa}R+\frac{2\kappa N}{\sqrt{q}}\left(\pi^{ab}\pi_{ab}-\frac{\pi^2}{2}\right)-\frac{\sqrt{q}}{\kappa}D_cD^cN+\pi D_cN^c+N^cD_c\pi\,.
\end{equation}
The equation found corresponds to eq.~\eqref{eq:limit-field-equation-u} when $\pi_u=-\pi$ in the limit $u=0$. Hence, eq.~\eqref{eq:limit-field-equation-u} is concluded to be the ADM-decomposed version of ${}^{(4)}R=0$, which is the limit of eq.~\eqref{eq:field-equation-u} for $u=0$, as expected.

\subsection{Projections of covariant field equations, constraints, and dynamical equations}
\label{eq:field-equations-constraints-relation}

Suitable projections of the Einstein equations along $n^{\mu}$ and into $\Sigma_t$ are known to reproduce the Hamiltonian and momentum constraints as well as the dynamical field equations; see refs.~\cite{Gourgoulhon:2007ue,Gourgoulhon:2012} for GR and ref.~\cite{Reyes:2022mvm} for a modified gravity governed by nondynamical background fields $u$ and $s^{\mu\nu}$ of the minimal gravitational SME. Here, our interest is to check whether these properties still hold in the setting of eq.~\eqref{eq:modified-einstein-equations}. To accomplish this endeavor, we decompose the covariant field equations into four contributions:
\begin{subequations}
\begin{equation}
\label{eq.Einstein_mod}
{}^{(4)}G_{\mu\nu}-{}^{(4)}T_{\mu\nu}^{(u,\omega,V)}=0\,,
\end{equation}
where we put all nonstandard terms depending on $u$ into the following sum of second-rank tensors,
\begin{equation}
{}^{(4)}T_{\mu\nu}^{(u,\omega,V)}=T_{\mu\nu}^{(u)}+T_{\mu\nu}^{(\omega)}+T_{\mu\nu}^{(V)}\,,
\label{eq:T(mu_nu)-geral}
\end{equation}
with
\begin{align}
\label{eq:T-mu_nu-u}
T_{\mu\nu}^{(u)}&=u{}^{(4)}G_{\mu\nu}-\nabla_\mu \nabla_\nu u+g_{\mu\nu}\square u\,, \\[1ex]
\label{eq:T-mu_nu-w}
T_{\mu\nu}^{(\omega)}&=\frac{\omega}{1-u}\bigg(\nabla_\mu u \nabla_\nu u-\frac{1}{2}g_{\mu\nu}\nabla_{\varrho}u\nabla^{\varrho} u\bigg)\,, \\[1ex]
\label{eq:T-mu_nu-U}
T_{\mu\nu}^{(V)}&=-\frac{1}{2}g_{\mu\nu}V(u)\,.
\end{align}
\end{subequations}
Now we intend to compute projections of the modified Einstein equations of eq.~\eqref{eq:modified-einstein-equations} onto the one-dimensional subspace orthogonal to $\Sigma_t$ and partially into $\Sigma_t$, respectively. Computational details are relegated to appendix~\ref{eq:projections-gravitational-field-equations}. Both projections are found to be
\begin{subequations}
\begin{align}
\label{eq:purely-orthogonal-projection-modified-einstein-equations}
0=2n^{\mu}n^{\nu}\Big({}^{(4)}G_{\mu\nu}-{}^{(4)}T_{\mu\nu}^{(u,\omega,V)}\Big)&=-\frac{2\kappa}{\sqrt{q}}\mathcal{H}\,, \\[2ex]
\label{eq:mixed-projection-modified-einstein-equations}
0=2n^{\mu}q^{\nu}_{\phantom{\nu}a}\Big({}^{(4)}G_{\mu\nu}-{}^{(4)}T_{\mu\nu}^{(u,\omega,V)}\Big)&=-\frac{2\kappa}{\sqrt{q}}\mathcal{H}_a\,,
\end{align}
\end{subequations}
with the Hamiltonian and momentum constraints $\mathcal{H}$ and $\mathcal{H}_a$ of eqs.~\eqref{eq:hamiltonian-constraint} and \eqref{eq:momentum-constraint}, respectively. Hence, the purely orthogonal projection of the modified Einstein equations is related to the Hamiltonian constraint, whereas the mixed projection provides the momentum constraints. This property is taken over directly from GR and is an independent crosscheck of the secondary constraints of eq.~\eqref{eq:u-action-ADM-decomposed}.

Moreover, working in coordinates where $N^a=0$, we are also able to show that
\begin{equation}
\label{eq:complete-projection-hypersurface}
0=q^a_{\phantom{a}\mu}q^b_{\phantom{b}\nu}\Big[{}^{(4)}G^{\mu\nu}-({}^{(4)}T^{(u,\omega,V)})^{\mu\nu}\Big]=\frac{2\kappa}{N\sqrt{q}}\Big(\dot{\pi}^{ab}-\{\pi^{ab}(x),H[N]\}\Big)\,,
\end{equation}
with the Poisson bracket of eq.~\eqref{eq:poisson-bracket-hamiltonian-constraint-pi}. Therefore, the modified Einstein equations completely projected into $\Sigma_t$ of the spacetime foliation reproduce the second set of Hamilton's equations for $q_{ab}$, which are the dynamical field equations for the gravitational field in the presence of $u$. All these demonstrations pose independent crosschecks of the constraints as well as the dynamics of the modified-gravity theory under study.

\section{Constraint algebra}
\label{sec:constraint-structure}

To be able to make a precise statement on the number of physical degrees of freedom in a modified-gravity theory, we must understand the time evolution of all the constraints.
An application of the Dirac-Bergmann algorithm \cite{Hanson:1976,Henneaux:1992} will provide the full number of constraints, i.e., it will also tell us whether additional ones arise beyond those of GR. We have already obtained the secondary constraints $\mathcal{H}\approx 0$ and $\mathcal{H}_a\approx 0$; see eqs.~\eqref{eq:secondary-constraints}. Thus, although the Hamiltonian in eq.~\eqref{eq:hamiltonian-scalar-model-after-integrations-by-parts} is modified as compared to that of GR, it still decomposes into parts multiplied by the lapse function and contracted with the shift vector components, respectively. So $\mathcal{H}\approx 0$ and $\mathcal{H}_a\approx 0$ can be interpreted as modified Hamiltonian and momentum constraints.

Now, the secondary constraints are subject to time evolution themselves, i.e., Poisson brackets must be computed with the augmented Hamiltonian $H_A$ of eq.~\eqref{eq:augmented-hamiltonian} and these constraints. Due to the structure of $H_A$ and
\begin{equation}
\{\pi_N,\mathcal{H}\}=\{\pi_N,\mathcal{H}_a\}=\{\pi_a,\mathcal{H}\}=\{\pi_a,\mathcal{H}_b\}=0\,,
\end{equation}
this amounts to computing Poisson brackets between the modified Hamiltonian and momentum constraints themselves. From a technical standpoint it is helpful to study the smeared constraints of eq.~\eqref{eq:constraints-smeared}. The computations are partially tedious and are relegated to appendices~\ref{sec:momentum-momentum-constraint} — \ref{sec:hamiltonian-momentum-constraint} for the reader interested in the details. The results can be expressed in a highly compact form:
\begin{subequations}
\label{eq:constraint-algebra}
\begin{align}
\label{eq:constraint-algebra-1}
\{P[\mathbf{N}],P[\mathbf{N}']\}&=P[(\mathbf{N}\cdot\mathbf{D})\mathbf{N}'-(\mathbf{N}'\cdot\mathbf{D})\mathbf{N}]\,, \\[2ex]
\label{eq:constraint-algebra-2}
\{H[N], H[N']\}&=P[N\mathbf{D}N'-N'\mathbf{D}N]\,, \\[2ex]
\label{eq:constraint-algebra-3}
\{P[\mathbf{N}],H[N]\}&=H[(\mathbf{N}\cdot\mathbf{D})N]\,.
\end{align}
\end{subequations}
Alternatively, by properly removing the smearing functions, which is to be demonstrated in appendix~\ref{sec:removal-smearing-functions}, the Poisson brackets between the constraints proper are put on paper as follows:
\begin{subequations}
\label{eq:constraint-algebra-without-smearing}
\begin{align}
\{H_a(x'),H_b(x'')\}&=\mathcal{H}_a(x')\partial_b\delta(x'-x'')-\mathcal{H}_b(x'')\partial_a\delta(x''-x')\,, \\[2ex]
\{H(x'),H(x'')\}&=\mathcal{H}_c(x')\partial^c\delta(x'-x'')-\mathcal{H}_c(x'')\partial^c\delta(x''-x')\,, \\[2ex]
\{P_a(x'),H(x'')\}&=\mathcal{H}(x')\partial_a\delta(x'-x'')\,.
\end{align}
\end{subequations}
Thus, the Hamiltonian and momentum constraints close under the Poisson bracket, i.e., the Poisson brackets of these constraints can again be expressed in terms of the constraints proper. The latter findings are remarkable and are a consequence of modifying GR by a dynamical scalar field instead of breaking diffeomorphism invariance explicitly. Preliminary computations were also performed for the $u$ sector considered in refs.~\cite{Reyes:2021cpx,Reyes:2022mvm,Reyes:2023sgk} with explicit diffeomorphism symmetry violation. Here, eq.~\eqref{eq:constraint-algebra} was not found to hold, which makes the behavior of such theories way more involved and obscure. So at the level of the Hamiltonian formulation, the secondary constraints of the dynamical $u$ sector is characterized by eq.~\eqref{eq:constraint-algebra}. These are taken directly over from GR where the usual Hamiltonian and momentum constraints are replaced by the modified constraints. Then, the time evolution of these constraints does not give rise to further constraints and the Dirac-Bergmann algorithm stops. So the augmented Hamiltonian of eq.~\eqref{eq:augmented-hamiltonian} is the full Hamiltonian of the modified-gravity theory under study; cf.~ref.~\cite{GabrieleGionti:2020drq}.

Having these results at hand, we can compute the number of physical gravitational degrees of freedom for the modified theory. Following ref.~\cite{Henneaux:1992}, the latter is given by
\begin{equation}
N_{\mathrm{dof}}=\frac{1}{2}(N_{\mathrm{ph}}-2N_1-N_2)\,,
\end{equation}
with the number of phase space variables $N_{\mathrm{ph}}$ and the numbers $N_1$ ($N_2$) of first-class (second-class) constraints. Now, since the Poisson brackets of all 8 constraints with each other do not provide new constraints, they are first-class. Second-class constraints are not found. As there are 10 independent metric components and 10 conjugated-momentum components, $N_{\mathrm{ph}}=20$. This counting amounts to 2 physical propagating gravitational degrees of freedom. As a consequence, the known number from GR is preserved as long as diffeomorphism invariance is violated spontaneously. As mentioned briefly before, eq.~\eqref{eq:constraint-algebra} breaks down when diffeomorphism invariance is violated explicitly. The number of first- and probably even second-class constraints is then expected to change, i.e., novel degrees of freedom emerge \cite{Kostelecky:2017zob,ONeal-Ault:2020ebv,Bailey:2024zgr}.

\section{Final remarks}
\label{sec:final-remarks}

One possibility of incorporating spacetime symmetry violation into the EH action is via a dynamical scalar field $u$ that acquires a nonvanishing position-dependent vacuum expectation value $\langle u\rangle$. As a result, some or all of the diffeomorphism generators are spontaneously broken depending on how $\langle u\rangle$ depends on the coordinates. Also, $e_a^{\phantom{a}\mu}\partial_{\mu}\langle u\rangle$ is capable of breaking a subset or all of the generators of the local $\mathit{SO}(3,1)$ group in freely falling inertial frames.

A characteristic of spontaneous spacetime symmetry violation at the level of the Hamiltonian formulation is that the Hamiltonian of the theory decomposes into four contributions that remain first-class constraints albeit modified by $u$. So the constraint structure of GR is actually preserved, whereupon there are 2 physical propagating degrees of freedom. Thus, a setting of spontaneous spacetime symmetry violation is capable of describing modifications of GR, while preserving at least some of its beautiful and valuable properties.

All of this is in stark contrast to explicit symmetry violation induced by the presence of a background field such as $u$, which is simply put into the EH action by hand and does not arise dynamically. Preliminary studies show that the Hamiltonian then does not decompose anymore into a set of first-class constraints, but the structure of constraints is heavily modified, potentially changing the number of physical gravitational degrees of freedom.

While explicit symmetry violation in gravity can be interesting in its own right, it has proven to be technically highly challenging to set up a meaningful Hamiltonian formulation for such a theory. Furthermore, the actual number of physical degrees of freedom is obscure in these cases. Thus, a viable alternative to such settings, which is way more tractable and nevertheless allows for exploring new, interesting physics beyond GR, is to violate its spacetime symmetries spontaneously. A proposal can be to focus on the study of the Hamiltonian formulation of such theories and assess possible implications on gravitational phenomenology.

\acknowledgments

It is a pleasure to thank R.S. Thibes for valuable discussions and comments on the results in the Hamiltonian formulation. M.S.
is indebted to FAPEMA Universal 00830/19, CNPq Produtividade 310076/2021-8, and CAPES/Finance Code 001.

\paragraph{Note added}

After completing this manuscript, we realized a connection between our analysis of the canonical field equations for eq.~\eqref{eq:u-action-ADM-decomposed} and that carried out in ref.~\cite{GiontiSJ:2023tgx}.

\appendix

\section{Field equations}
\label{sec:field-equations-derivations}

In the following, we would like to provide some details on how to derive the covariant field equations for eq.~\eqref{eq:action-BD-type-theory} with $\phi=1-u$.

\subsection{Gravitational field}
\label{sec:field-equations-gravitational-field}

Varying the first term of eq.~\eqref{eq:action-BD-type-theory}, with $\phi=1-u$ understood, leads to
\begin{align}
\delta\left(\sqrt{-g}(1-u){}^{(4)}R\right)&=\delta(\sqrt{-g})(1-u){}^{(4)}R+\sqrt{-g}\delta\left[(1-u){}^{(4)}R\right] \notag \\
&=-\frac{1}{2}\sqrt{-g}g_{\mu\nu}\delta g^{\mu\nu}(1-u){}^{(4)}R+\sqrt{-g}(1-u)\delta {}^{(4)}R\,.
\end{align}
We need the variation of the Ricci scalar, which is suitably expressed in terms of the variation of the Ricci tensor as follows:
\begin{align}
\delta\left(\sqrt{-g}(1-u){}^{(4)}R\right)&=\sqrt{-g}\delta g^{\mu\nu}(1-u)\bigg({}^{(4)}R_{\mu\nu}-\frac{1}{2}g_{\mu\nu}{}^{(4)}R\bigg) \notag \\
&\phantom{{}={}}+\sqrt{-g}(g^{\mu\nu}\delta {}^{(4)}R_{\mu\nu}-ug^{\mu\nu}\delta {}^{(4)}R_{\mu\nu})\,,
\end{align}
which allows us to employ the contracted Palatini identity
\begin{equation}
g^{\mu\nu}\delta {}^{(4)}R_{\mu\nu}=(\nabla^\mu \nabla^\nu-g^{\mu\nu}\square)\delta g_{\mu\nu}\,.
\end{equation}
By taking into account the extended GHY boundary term of eq.~\eqref{eq:action-BD-type-theory}, integrations by parts provide
\begin{equation}
-ug^{\mu\nu}\delta {}^{(4)}R_{\mu\nu}=(\nabla^\mu \nabla^\nu u-g_{\mu\nu}\square u)\delta g^{\mu\nu}\,.
\end{equation}
whereupon we arrive at
\begin{equation}
\delta\left(\sqrt{-g}(1-u){}^{(4)}R\right)=\sqrt{-g}\delta g^{\mu\nu}\left[(1-u){}^{(4)}G_{\mu\nu}+(\nabla^\mu \nabla^\nu u-g_{\mu\nu}\square u)\right]\,.
\label{eq:termo_1}
\end{equation}
For the second term of eq.~\eqref{eq:action-BD-type-theory} we have
\begin{align}
    \delta\bigg(-\sqrt{-g}\frac{\omega}{1-u}g^{\mu\nu}\nabla_\mu u\nabla_\nu u\bigg)&=-\delta(\sqrt{-g})\frac{\omega}{1-u}g^{\mu\nu}\nabla_\mu u \nabla_\nu u-\sqrt{-g}\delta g^{\mu\nu}\frac{\omega}{1- u}\nabla_\mu u \nabla_\nu u \notag \\
&=-\sqrt{-g}\delta g^{\mu\nu}\frac{\omega}{1-u}\bigg(\nabla_\mu u \nabla_\nu u-\frac{1}{2}g_{\mu\nu}\nabla_\mu u \nabla^\mu u\bigg)\,.\label{eq:termo_2}
\end{align}
Finally, varying the third term of eq.~\eqref{eq:action-BD-type-theory} implies
\begin{equation}
\delta\Big(-\sqrt{-g}V(u)\Big)=\sqrt{-g}\delta g^{\mu\nu}\frac{1}{2}g_{\mu\nu}V(u)\,.
\label{eq:termo_3}
\end{equation}
Inserting eqs.~\eqref{eq:termo_1}, \eqref{eq:termo_2}, and \eqref{eq:termo_3} into \eqref{eq:action-BD-type-theory} results in
\begin{align}
    \delta(S_{\mathrm{ST}})&=\frac{1}{2\kappa}\int_{\mathcal{M}} \mathrm{d}^4x\,\sqrt{-g}\delta g^{\mu\nu}\bigg[(1-u){}^{(4)}G_{\mu\nu}+\left(\nabla^\mu \nabla^\nu u-g_{\mu\nu}\square u\right)\notag \\
    &\phantom{{}={}}\hspace{3.7cm}-\frac{\omega}{1-u}\bigg(\nabla_\mu u \nabla_\nu u-\frac{1}{2}g_{\mu\nu}\nabla_\mu u \nabla^\mu u\bigg)+\frac{1}{2}g_{\mu\nu}V(u)\bigg]\,.
\end{align}
Then, the variational principle leads to the field equations given by eq.~\eqref{eq:modified-einstein-equations}.

\subsection{Dynamical scalar field}
\label{sec:field-equations-u}

Next, we are interested in obtaining the field equations for the dynamical scalar field $u$. From the first term of eq.~\eqref{eq:action-BD-type-theory} we have 
\begin{align}
\delta_u\bigg(\frac{\omega}{1-u}g^{\mu\nu}\nabla_{\mu}u\nabla_{\nu}u\bigg)&=\frac{\omega}{(1-u)^2}g^{\mu\nu}\nabla_{\mu}u\nabla_{\nu}u+\frac{\omega}{1-u}g^{\mu\nu}\delta_u\left(\nabla_{\mu} u\nabla_{\nu}u\right) \notag \\
&=-\frac{\omega}{1-u}\bigg(\frac{1}{1-u}\nabla_{\mu}u\nabla^{\mu}u+2\square u\bigg)\delta u\,,
\end{align}
such that
\begin{equation}
\delta_uS_{\mathrm{ST}}=\frac{1}{2\kappa}\int_{\mathcal{M}} \mathrm{d}^4x\,\sqrt{-g}\left[-{}^{(4)}R+\frac{\omega}{1-u}\left(\frac{1}{1-u}\nabla_{\mu}u\nabla^{\mu}u+2\square u\right)-\frac{\delta V(u)}{\delta u}\right]\delta u\,,
\end{equation}
where the variation of the potential for $u$ can simply be interpreted as the first derivative $V'(u)$. This provides the field equation for $u$ of eq.~\eqref{eq:field-equation-u}.

\subsection{Projections of gravitational field equations}
\label{eq:projections-gravitational-field-equations}

In the following, we will show more details on how to compute various projections of the modified gravitational field equations in their covariant form; see eq.~\eqref{eq:modified-einstein-equations}. We shall find that particular projections provide the Hamiltonian and momentum constraints as well as the dynamical part of the field equations.

\subsubsection{Purely orthogonal projection}
\label{eq:modified-einstein-equations-purely-orthogonal}

To start, we compute the projection completely orthogonal to $\Sigma_t$, i.e., we contract twice with~$n^{\mu}$:
\begin{equation}
\label{eq:proj_ortogonal}
2n^{\mu}n^{\nu}\Big({}^{(4)}G_{\mu\nu}-{}^{(4)}T_{\mu\nu}^{(u,\omega,V)}\Big)=0\,,
\end{equation}
where the second part is given by eq.~\eqref{eq:T(mu_nu)-geral}.
The projections of the Einstein tensor and $T_{\mu\nu}^{(u)}$ were already obtained in ref.~\cite{Reyes:2021cpx} with great detail. Thus, we are going to quote these results without derivation:
\begin{subequations}
\begin{equation}
2n^{\mu}n^{\nu}\Big({}^{(4)}G_{\mu\nu}-T_{\mu\nu}^{(u)}\Big)=(1-u)\big(R+K^2-K_{ab}K^{ab}\big)+2D_aD^au-\frac{2}{N}K\mathcal{L}_m u\,,
\end{equation}
where
\begin{align}
\label{tr-K}
K&=-\frac{1}{2(1-u)}\left(2\kappa\frac{\pi}{\sqrt{q}}-\frac{3}{N}\mathcal{L}_m u\right)\,, \\[1ex]
K^{ab}K_{ab}&=\frac{1}{(1-u)^2}\left[\frac{4\kappa^2}{q}\left(\pi^{ab}\pi_{ab}-\frac{\pi^2}{4}\right)-\frac{2\kappa}{\sqrt{q}}\frac{\pi}{2N}\mathcal{L}_m u+\frac{3}{4N^2}(\mathcal{L}_m u)^2\right]\,,
\end{align}
\end{subequations}
and $\mathcal{L}_mu$ given in eq.~\eqref{eq:poisson-bracket-hamiltonian-constraint-u}. Inserting all these results properly gives rise to
\begin{align}
\label{eq:proj_G-T(u)}
2n^{\mu}n^{\nu}\Big({}^{(4)}G_{\mu\nu}-T_{\mu\nu}^{(u)}\Big)&=(1-u)R-\frac{4\kappa^2}{(1-u)q}\left(\pi_{ab}\pi^{ab}-\frac{\pi^2}{2}\right) \notag \\
&\phantom{{}={}}-\frac{3}{2}\frac{4\kappa^2}{(1-u)q}\frac{1}{(3+2\omega)^2}[\pi+(1-u)\pi_u]^2+2D_aD^au\,.
\end{align}
Next, we need the projection of \eqref{eq:T-mu_nu-w}:
\begin{align}
2n^{\mu}n^{\nu}T_{\mu\nu}^{(\omega)}&=\frac{\omega}{1-u}\bigg(2n^{\mu}n^{\nu}\nabla_\mu u \nabla_\nu u-n^{\mu}n^{\nu}g_{\mu\nu}\nabla_{\varrho} u \nabla^{\varrho} u\bigg) \notag \\
&=\frac{\omega}{1-u}\bigg(2n^{\mu}n^{\nu}\nabla_\mu u \nabla_\nu u+\left(q^{\mu\nu}-n^{\mu}n^{\nu}\right)\nabla_\mu u \nabla_\nu u\bigg)\,,
\end{align}
where we employ that $n^2=-1$ as well as the completeness relation $g^{\mu\nu}=q^{\mu\nu}-n^{\mu}n^{\nu}$. We then benefit from
\begin{subequations}
\begin{align}
n^{\mu}n^{\nu}\nabla_\mu u \nabla_\nu u=\frac{1}{N^2}\big(\mathcal{L}_{m}u\big)^2\,, \\[1ex]
q^{\mu\nu}\nabla_\mu u \nabla_\nu u=q^{ab}D_a u D_b u\,,
\end{align}
\end{subequations}
which implies
\begin{equation} 2n^{\mu}n^{\nu}T_{\mu\nu}^{(\omega)}=\frac{\omega}{1-u}\bigg(\frac{4\kappa^2}{q}\frac{1}{(3+2\omega)^2}[\pi+(1-u)\pi_u]^2+D_a u D^a u\bigg)\,. \label{eq:proj_T(w)}
\end{equation}
The final projection of eq.~\eqref{eq:T-mu_nu-U} is quickly obtained as
\begin{equation}
\label{eq:proj_T(U)}
2n^{\mu}n^{\nu}T_{\mu\nu}^{(V)}=-n^{\mu}n^{\nu}g_{\mu\nu}V(u)=-n^{\mu}n_{\mu}V(u)=V(u)\,.
\end{equation}
By inserting eqs.~\eqref{eq:proj_G-T(u)}, \eqref{eq:proj_T(w)} and \eqref{eq:proj_T(U)} into eq.~\eqref{eq:proj_ortogonal}, we arrive at eq.~\eqref{eq:purely-orthogonal-projection-modified-einstein-equations}.

\subsubsection{Mixed projection}
\label{eq:modified-einstein-equations-mixed}

Here, we compute the partial projection of the covariant gravitational field equations into $\Sigma_t$, i.e., the latter is to be contracted with the orthogonal direction $n^{\mu}$ and the projector $q^{\nu}_{\phantom{\nu}a}$ into $\Sigma_t$:
\begin{equation}
2n^{\mu}q^{\nu}_{\phantom{\nu}a}\Big({}^{(4)}G_{\mu\nu}-{}^{(4)}T_{\mu\nu}^{(u,\omega,V)}\Big)=0\,,
    \label{eq:proj_parcial}
\end{equation}
with the second part stated in eq.~\eqref{eq:T(mu_nu)-geral}. The first term is evaluated by means of the contracted Codazzi-Mainardi equation:
\begin{equation}
2n^{\mu}q^{\nu}_{\phantom{\nu}a}{}^{(4)}G_{\mu\nu}=2(D^c K_{ca}-D_aK)\,.
    \label{proj_parcial_G(mu,nu)}
\end{equation}
As before, we will evaluate the projections of the three tensors contained in ${}^{(4)}T_{\mu\nu}^{(u,\omega,V)}$ step by step. The first term of eq.~\eqref{eq:T-mu_nu-u} has the form of eq.~\eqref{proj_parcial_G(mu,nu)} multiplied with the scalar field $u$. Therefore,
\begin{equation}
2n^{\mu}q^{\nu}_{\phantom{\nu}a}u{}^{(4)}G_{\mu\nu}=2u(D^c K_{ca}-D_aK)\,.
    \label{proj_parcial_G(mu,nu)_u}
\end{equation}
To evaluate the projection of the second term, we benefit from a valuable relationship analogous to eq.~(F38) in ref.~\cite{Reyes:2021cpx}:
\begin{align}
-2n^{\mu}q^{\nu}_{\phantom{\nu}a}\nabla_\mu \nabla_\nu u&=-2D_a\bigg(\frac{1}{N}\mathcal{L}_m u\bigg)+2K_{ca}D^cu\,.
    \label{proj_parcial_termo2_T(u)}
\end{align}
Considering the identity $n^\mu q^{\nu}_{\phantom{{\nu}}\mu}=0$, it is clear that the terms proportional to $g_{\mu\nu}$ in eqs.~\eqref{eq:T-mu_nu-u}, \eqref{eq:T-mu_nu-w}, and \eqref{eq:T-mu_nu-U} do not contribute. For the final term, i.e., the first contribution of eq.~\eqref{eq:T-mu_nu-w} we have
\begin{equation}
2n^{\mu}q^{\nu}_{\phantom{\nu}a}T^{(\omega)}_{\mu\nu}=\frac{\omega}{1-u}\bigg(2n^{\mu}q^{\nu}_{\phantom{\nu}a}\nabla_\mu u \nabla_\nu u\bigg)=\frac{\omega}{1-u}\bigg(\frac{2}{N}\mathcal{L}_m u D_a u\bigg)\,.
    \label{proj_parcial_T(w)}
\end{equation}
Then, summing eqs.~\eqref{proj_parcial_G(mu,nu)}, \eqref{proj_parcial_G(mu,nu)_u}, \eqref{proj_parcial_termo2_T(u)}, and \eqref{proj_parcial_T(w)} and inserting these into eq.~\eqref{eq:proj_parcial} implies:
\begin{align}
2n^{\mu}q^{\nu}_{\phantom{\nu}a}\Big({}^{(4)}G_{\mu\nu}-{}^{(4)}T_{\mu\nu}^{(u,\omega,V)}\Big)&=2(1-u)(D^cK_{ca}-D_aK)-2K_{ca}D^cu \notag \\[1ex]
&\phantom{{}={}}+2D_a\bigg(\frac{1}{N}\mathcal{L}_m u\bigg)-\frac{2\omega}{1-u}\bigg(\frac{1}{N}\mathcal{L}_m u D_a u\bigg)\,,
    \label{eq:proj_parcial_2}
\end{align}
where $K$ and $K_{ca}$ can be taken from eqs.~\eqref{eq:extrinsic-curvature-scalar-canonical-momentum}, \eqref{eq:extrinsic-curvature-tensor-canonical-momentum}. For the first term on the right-hand side of eq.~\eqref{eq:proj_parcial_2} we have
\begin{align}
    D^c K_{ca}-D_aK&=\frac{D^cu}{(1-u)^2}\left[\frac{2\kappa}{\sqrt{q}}\left(\pi_{ca}-\frac{\pi}{2}q_{ca}\right)+\frac{q_{ca}}{2N}\mathcal{L}_m u\right] \notag \\
    &\phantom{{}={}}+\frac{1}{(1-u)}\left[\frac{2\kappa}{\sqrt{q}}D^c\left(\pi_{ca}-\frac{\pi}{2}q_{ca}\right)+\frac{q_{ca}}{2N}D^c(\mathcal{L}_m u)\right] \notag \\
    &\phantom{{}={}}+\frac{D_au}{2(1-u)^2}\left(2\kappa\frac{\pi}{\sqrt{q}}-\frac{3}{N}\mathcal{L}_m u\right)+\frac{1}{2(1-u)}D_a\left(2\kappa\frac{\pi}{\sqrt{q}}-\frac{3}{N}\mathcal{L}_m u\right) \notag\\
    &=\frac{1}{(1-u)}\bigg[K_{ca}D^cu+\frac{2\kappa}{\sqrt{q}}\bigg(D^c\pi_{ca}-\frac{1}{2}D_a\pi\bigg) \notag \\
    &\phantom{{}={}}+\frac{1}{2}D_a\bigg(\frac{1}{N}\mathcal{L}_mu\bigg)-KD_au+\frac{2\kappa}{\sqrt{q}}\frac{1}{2}D_a\pi-\frac{3}{2}D_a\bigg(\frac{1}{N}\mathcal{L}_mu\bigg)\bigg]\,,
\end{align}
which, by rearranging the individual contributions, further leads to
\begin{align}
2(1-u)(D^c K_{ca}-D_aK)&=2(K_{ca}D^cu-KD_au)-2D_a\bigg(\frac{1}{N}\mathcal{L}_mu\bigg)+\frac{4\kappa}{\sqrt{q}}D^c\pi_{ca}\,.
\label{eq:proj_parcial_termo1}
\end{align}
Note that the second and third terms of eq.~\eqref{eq:proj_parcial_2} cancel each other. Hence, we rewrite the mixed projection of the gravitational field equations as
\begin{align}
2n^{\mu}q^{\nu}_{\phantom{\nu}a}\Big({}^{(4)}G_{\mu\nu}-{}^{(4)}T_{\mu\nu}^{(u,\omega,V)}\Big)&=\frac{4\kappa}{\sqrt{q}}D^c\pi_{ca}-2KD_au-\frac{2\omega}{1-u}\bigg(\frac{1}{N}\mathcal{L}_m u D_a u\bigg)\,.
    \end{align}
Summing the last two terms of the latter results in 
\begin{align}
-2KD_au-\frac{2\omega}{1-u}\bigg(\frac{1}{N}\mathcal{L}_m u D_a u\bigg)&=\frac{1}{1-u}\frac{2\kappa}{\sqrt{q}}\pi D_au-\frac{3+2\omega}{(1-u)N}\mathcal{L}_mu D_au \notag \\
&=\frac{1}{1-u}\frac{2\kappa}{\sqrt{q}}\pi D_au \notag \\
&\phantom{{}={}}-\frac{3+2\omega}{(1-u)N}\bigg(\frac{2\kappa N}{\sqrt{q}}\frac{1}{3+2\omega}\big[\pi+(1-u)\pi_u\big]\bigg)D_au\notag \\
&=\frac{1}{1-u}\frac{2\kappa}{\sqrt{q}}\pi D_au-\frac{1}{1-u}\frac{2\kappa}{\sqrt{q}}\pi D_au-\frac{2\kappa}{\sqrt{q}}\pi_u D_au\notag\\
&=-\frac{2\kappa}{\sqrt{q}}\pi_u D_au\,.
\label{eq:proj_parcial_termo2}
\end{align}
Finally, we conclude that
\begin{equation}
2n^{\mu}q^{\nu}_{\phantom{\nu}a}\Big({}^{(4)}G_{\mu\nu}-{}^{(4)}T_{\mu\nu}^{(u,\omega,V)}\Big)=\frac{4\kappa}{\sqrt{q}}D_c\pi^c_{\phantom{c}a}-\frac{2\kappa}{\sqrt{q}}\pi_u D_au\,,
\end{equation}
which provides eq.~\eqref{eq:mixed-projection-modified-einstein-equations}.

\subsubsection{Complete projection into the hypersurface}

Ultimately, we intend to project the gravitational field equations completely into $\Sigma_t$. This means that we must contract the latter with two projectors $q^a_{\phantom{i}\mu}$. Doing so for the combination of the Einstein tensor and $(T^{(u)})^{\mu\nu}$ of eq.~\eqref{eq:T-mu_nu-u} provides a known result (see ref.~\cite{Reyes:2022mvm}):
\begin{align}
\label{eq:spacelike-projection-1-2}
\frac{N\sqrt{q}}{2\kappa}q^a_{\phantom{a}\mu}q^b_{\phantom{b}\nu}\Big[{}^{(4)}G^{\mu\nu}-(T^{(u)})^{\mu\nu}\Big]&=\dot{\pi}^{ab}+(1-u)\frac{N\sqrt{q}}{2\kappa}\left(R^{ab}-\frac{R}{2}q^{ab}\right) \notag \\
&\phantom{{}={}}+\frac{\sqrt{q}}{2\kappa}\left\{q^{ab}D_cD^c[(1-u)N]-D^aD^b[(1-u)N]\right\} \notag \\
&\phantom{{}={}}+\frac{\sqrt{q}}{2\kappa}\left[q^{ab}D^cND_cu-(D^aND^bu+D^bND^au)\right] \notag \\
&\phantom{{}={}}+\frac{2\kappa N}{\sqrt{q}(1-u)}\left[2\pi^{ac}\pi_c^{\phantom{c}b}-\pi\pi^{ab}-\frac{1}{2}\left(\pi_{cd}\pi^{cd}-\frac{\pi^2}{2}\right)q^{ab}\right] \notag \\
&\phantom{{}={}}+\frac{\mathcal{L}_mu}{1-u}\left(\pi^{ab}-\frac{3}{8}\frac{\sqrt{q}}{\kappa N}\mathcal{L}_mq^{ab}\right)\,.
\end{align}
We must now express the Lie derivative of the scalar field in terms of the canonical momenta according to eq.~\eqref{eq:poisson-bracket-hamiltonian-constraint-u}. In particular, for the last two lines of the latter finding:
\begin{align}
\label{eq:spacelike-projection-3}
\frac{\mathcal{L}_mu}{1-u}\left(\pi^{ab}-\frac{3}{8}\frac{\sqrt{q}}{\kappa N}\mathcal{L}_mq^{ab}\right)&=\frac{2\kappa N}{\sqrt{q}(1-u)}\frac{1}{3+2\omega}\big[\pi+(1-u)\pi_u\big]\pi^{ab} \notag \\
&\phantom{{}={}}-\frac{3}{2}\frac{\kappa N}{\sqrt{q}(1-u)}\frac{1}{(3+2\omega)^2}\Big[\pi+(1-u)\pi_u\Big]^2q^{ab}\,.
\end{align}
Performing the projection for the third term of eq.~\eqref{eq:T-mu_nu-w} gives
\begin{align}
\label{eq:spacelike-projection-4}
\frac{N\sqrt{q}}{2\kappa}q^a_{\phantom{a}\mu}q^b_{\phantom{b}\nu}(T^{(\omega)})^{\mu\nu}&=\frac{\omega}{1-u}\left(q^a_{\phantom{a}\mu}q^b_{\phantom{b}\nu}\nabla^{\mu}u\nabla^{\nu}u-\frac{1}{2}q^a_{\phantom{a}\mu}q^b_{\phantom{b}\nu}g^{\mu\nu}\nabla_{\varrho}u\nabla^{\varrho}u\right) \notag \\
&=\frac{N\sqrt{q}}{2\kappa}\frac{\omega}{1-u}\left[D^auD^bu-\frac{1}{2}q^{ab}(q_{\varrho\sigma}-n_{\varrho}n_{\sigma})\nabla^{\varrho}u\nabla^{\sigma}u\right] \notag \\
&=\frac{N\sqrt{q}}{2\kappa}\frac{\omega}{1-u}\left(D^auD^bu-\frac{1}{2}q^{ab}D_cuD^cu+\frac{1}{2N^2}q^{ab}(\mathcal{L}_mu)^2\right) \notag \\
&=\frac{N\sqrt{q}}{2\kappa}\frac{\omega}{1-u}\left(D^auD^bu-\frac{1}{2}q^{ab}D_cuD^cu\right) \notag \\
&\phantom{{}={}}+\frac{N\kappa}{\sqrt{q}(1-u)}\frac{\omega}{(3+2\omega)^2}\Big[\pi+(1-u)\pi_u\Big]^2q^{ab}\,.
\end{align}
The projection of the fourth term of eq.~\eqref{eq:T-mu_nu-U} leads to
\begin{equation}
q^a_{\phantom{a}\mu}q^b_{\phantom{b}\nu}(T^{(V)})^{\mu\nu}=-\frac{1}{2}q^{ab}V(u)\,.
\end{equation}
Now, by summing eqs.~\eqref{eq:spacelike-projection-1-2}, \eqref{eq:spacelike-projection-3}, and \eqref{eq:spacelike-projection-4} we arrive at eq.~\eqref{eq:complete-projection-hypersurface}.

\section{Demonstration of constraint algebra}
\label{eq:constraint-algebra-computation}

To avoid the presentation of lengthy calculations in the main body of the paper, we provide here the main necessary steps to derive the constraint algebra of eq.~\eqref{eq:constraint-algebra}.

\subsection{Momentum with momentum constraint}
\label{sec:momentum-momentum-constraint}

We must evaluate the Poisson bracket of smeared momentum constraints:
\begin{equation}
\{P[\mathbf{N}], P[\mathbf{N'}]\}=\int_{\Sigma_t} \mathrm{d}^3z\,\bigg(\frac{\delta P[\mathbf{N}]}{\delta q_{ab}(z)}\frac{\delta P[\mathbf{N}']}{\delta \pi^{ab}(z)}+\frac{\delta P[\mathbf{N}]}{\delta u(z)}\frac{\delta P[\mathbf{N}']}{\delta \pi_u(z)}-(\mathbf{N}\leftrightarrow \mathbf{N}')\bigg)\,,
\end{equation}
with $P[\mathbf{N}]$ stated in eq.~\eqref{eq:momentum-constraint-smeared}, where eq.~\eqref{eq:momentum-constraint} is used. The momentum constraint decomposes into a sum of the GR momentum constraint and a nonstandard contribution. Therefore, each part can be treated independently of the other. For brevity, the dependence of each field on $z$ will be suppressed. For the modified term we use that
\begin{subequations}
\begin{align}
    \frac{\delta P[\mathbf{N}]}{\delta u}&=-(\pi_uD_a N^a+N^aD_a\pi_u)=-\mathcal{L}_{\mathbf{N}}\pi_u\,, \\[2ex]
     \frac{\delta P[\mathbf{N}]}{\delta \pi_u}&=N^aD_au=\mathcal{L}_{\mathbf{N}}u\,.
\end{align}
\end{subequations}
Then,
\begin{align}
\frac{\delta P[\mathbf{N}]}{\delta u(z)}\frac{\delta P[\mathbf{N}']}{\delta \pi_u(z)}-(\mathbf{N}\leftrightarrow \mathbf{N}')&=-(\pi_uD_aN^a+N^aD_a\pi_u)N'^{b}D_bu \notag \\
&\phantom{{}={}}+N^aD_au(\pi_uD_bN'^{b}+N'^{b}D_b\pi_u) \notag \\
&=-\pi_uD_aN^{a}N'^{b}D_bu-N^aD_a\pi_uN'^{b}D_bu \notag \\
&\phantom{{}={}}+N^aD_au\pi_uD_bN'^{b}+N^aD_auN'^{b}D_b\pi_u \notag \\
&\simeq-N'^{b}D_aN^{a}\pi_uD_bu+(N'^{b}D_iN^{a}\pi_uD_bu+N^aD_iN'^b\pi_uD_bu \notag \\
&\phantom{{}={}}+N^aN'^b\pi_uD_aD_bu)+N^aD_bN'^{b}\pi_uD_au-(N'^{b}D_bN^a\pi_uD_au \notag \\
&\phantom{{}={}}+N^aD_bN'^b\pi_uD_au+N^aN'^b\pi_u D_bD_au) \notag \\
&=N^aD_aN'^b\pi_uD_bu-N'^{b}D_bN^a\pi_uD_au \notag \\
&=(N^bD_jN'^a-N'^{b}D_jN^a)\pi_uD_au\,,
\end{align}
which was reformulated by using the following relations:
\begin{subequations}
\begin{align}
D_a(N^aN'^b\pi_uD_bu)&=N'^{b}D_aN^{a}\pi_uD_bu+N^aD_aN'^b\pi_uD_bu \notag \\
&\phantom{{}={}}+N^aD_a\pi_uN'^{b}D_bu+N^aN'^b\pi_uD_aD_bu\,, \\[2ex]
D_b(N^aN'^b\pi_uD_au)&=N'^{b}D_bN^a\pi_uD_au+N^aD_bN'^b\pi_uD_au \notag \\
&\phantom{{}={}}+N^aD_auN'^{b}D_b\pi_u+N^aN'^b\pi_u D_bD_au\,.
\end{align}
\end{subequations}
We use the symbol `$\simeq$' to indicate that total covariant derivatives are dropped.
So we conclude that
\begin{align}
\label{eq:momentum-momentum-smeared}
\{P[\mathbf{N}],P[\mathbf{N}']\}&=\int_{\Sigma_t}\mathrm{d}^3z\,(N^bD_bN'^a-N'^bD_bN^a)(\pi_uD_au-2D_b\pi_a^{\phantom{a}b}) \notag \\
&=P[(\mathbf{N}\cdot\mathbf{D})\mathbf{N}'-(\mathbf{N}'\cdot\mathbf{D})\mathbf{N}]\,.
\end{align}
The latter result corresponds to eq.~\eqref{eq:constraint-algebra-1}.

\subsection{Hamiltonian with Hamiltonian constraint}
\label{sec:hamiltonian-hamiltonian-constraint}

Next, we ought to calculate the Poisson bracket of smeared Hamiltonian constraints:
\begin{equation}
\label{PB_Hamiltoniano}
\{H[N], H[N']\}=\int_{\Sigma_t} \mathrm{d}^3z\bigg[\frac{\delta H[N]}{\delta q_{ab}(z)}\frac{\delta H[N']}{\delta \pi^{ab}(z)}+\frac{\delta H[N]}{\delta u(z)}\frac{\delta H[N']}{\delta \pi_u(z)}-(N\leftrightarrow N')\bigg]\,,
\end{equation}
where $H[N]$ is given by eq.~\eqref{eq:hamiltonian-constraint-smeared}.
Note that $q_{ab}$ and $u$ occur in the Hamilton density both in an algebraic and a nonalgebraic manner where the canonical momenta $\pi^{ab}$ and $\pi_u$ can only be found in the algebraic contributions. This implies that the completely algebraic terms of the variation do not contribute to the Poisson brackets, but only the nonalgebraic contributions being multiplied with the algebraic terms of their respective canonical momenta.

Nonalgebraic contributions of variations for $q_{ij}$ are found in the terms $-N\sqrt{q}(1-u)R$ and $-2N\sqrt{q}D^aD_au$. The following results are helpful:
\begin{subequations}
\begin{align}
\label{eq:variation-determinant-q}
\delta\sqrt{q}&=\frac{\sqrt{q}}{2}q^{ab}\delta q_{ab}\,, \\[2ex]
\label{eq:variation-inverse-q}
\delta q^{ab}&=-q^{ac}q^{bd}\delta q_{cd} \,, \\[2ex]
\label{varconexaohipersup}
\delta\Gamma_{\phantom{\rho}\mu\nu}^{\rho}&=\frac{1}{2}q^{\rho\beta}\left( D_{\mu }\delta q_{\nu\beta}+D_{\nu}\delta q_{\mu\beta}-D_{\beta}\delta q_{\mu\nu}\right)\,, \\[2ex]
\label{varconexaohipersup2}
\delta \Gamma^{\mu}_{\phantom{\mu}\mu\nu}&=\frac{1}{2}q^{\beta\mu}D_\nu \delta q_{\beta\mu}\,.
\end{align}
\end{subequations}
Let us compute the variation of $-N\sqrt{q}(1-u)R$:
\begin{equation}
\delta(-N(1-u)\sqrt{q}R)=-N(1-u)[R \delta \sqrt{q}+\sqrt{q}\delta R]\,.
\end{equation}
To evaluate the first term, we employ eq.~\eqref{eq:variation-determinant-q}. The second contribution follows from
\begin{equation}
\sqrt{q}\delta(q^{ab}R_{ab})=\sqrt{q}(R_{ab}\delta q^{ab}+q^{ab}\delta R_{ab})\,.
\end{equation}
We now benefit from eq.~\eqref{eq:variation-inverse-q} such that
\begin{equation}
\sqrt{q}R_{ab}\delta q^{ab}=-\sqrt{q}R^{ab}\delta q_{ab}\,,
\end{equation}
where for the other term we consider the explicit form of the Ricci tensor in terms of the variation of the Christoffel symbol. Using metric compatibility, $D_\lambda q_{ab}=0$, we have
\begin{align}
\sqrt{q}q^{ab}\delta R_{ab}=&\sqrt{q}D_k(q^{ab}\delta \Gamma^k_{\phantom{k}ab})-\sqrt{q}D_a(q^{ab}\delta \Gamma^k_{\phantom{k}kb})=\sqrt{q}D_k(q^{ab}\delta \Gamma^k_{\phantom{k}ab}-q^{ak}\delta \Gamma^b_{\phantom{b}ba})\,.
\end{align}
Then, from the definitions,
\begin{align}
N\sqrt{q}q^{ab}\delta R_{ab}&=N\sqrt{q}D_k\bigg[\frac{1}{2}q^{ab}q^{kc}(D_a\delta q_{bc}+D_b \delta q_{ac}-D_c \delta q_{ab})-\frac{1}{2}q^{ak}q^{cb}D_a \delta q_{cb}\bigg] \notag \\
&=N\sqrt{q}D_k\bigg[\frac{1}{2}q^{ab}q^{kc}(2D_a\delta q_{bc})-\frac{1}{2}q^{cb}q^{ka}(2D_a\delta q_{cb})\bigg] \notag \\
&=N\sqrt{q}D_k\bigg(q^{ab}q^{kc}-q^{cb}q^{ka}\bigg)D_a\delta q_{cb} \notag \\
&=N\sqrt{q}\bigg(D^cD^b\delta q_{cb}-q^{cb}D^aD_a\delta q_{cb}\bigg)\,.
\end{align}
We also employ that
\begin{subequations}
\begin{align}
D^c(\sqrt{q}ND^b\delta q_{cb})-D^b(\sqrt{q}D^cN\delta q_{cb})&=\sqrt{q}\Big(ND^cD^b\delta q_{cb}-D^bD^cN\delta q_{cb}\Big)\,, \\[2ex]
D^a(\sqrt{q}Nq^{cb}D_a\delta q_{cb})-D^a(\sqrt{q}D_aNq^{cb}\delta q_{cb})&=\sqrt{q}q^{cb}\Big(ND^aD_a\delta q_{cb}-D^aD_aN\delta q_{cb}\Big)\,,
\end{align}
\end{subequations}
which leads to
\begin{equation}
N\sqrt{q}q^{ab}\delta R_{ab}\simeq\sqrt{q}\bigg[(D^c D^b N)\delta q_{cb}-q^{cb}(D^a D_a N)\delta q_{cb})\bigg]\,.
\end{equation}
Hence,
\begin{equation}
\label{vartermo1}
-(1-u)\frac{\delta (N\sqrt{q} R)}{\delta q_{ab}}\simeq \sqrt{q}\bigg((1-u)NG^{ab}+q^{ab}D^c D_c [(1-u)N]-D^a D^b [(1-u)N]\bigg)\,.
\end{equation}
Note that in the computation of the Poisson bracket, the term involving the Einstein tensor $G^{ij}$ will cancel, as it provides a contribution symmetric with respect to $N\leftrightarrow N'$.

Next, let us determine the variation of $-2N\sqrt{q}D^aD_au$. Here, we need to use eqs.~\eqref{eq:variation-determinant-q} and \eqref{eq:variation-inverse-q}:
\begin{align}
\delta(-2N\sqrt{q}D^aD_au)&=\delta\sqrt{q}(2 D^aND_au)+2\sqrt{q}\delta(q^{ab}D_bND_au) \notag \\
&=\frac{\sqrt{q}}{2}q^{ab}\delta q_{ab}(2D^aND_au)-2\sqrt{q}\delta q_{cd}q^{ac}q^{bd}D_bND_au \notag \\
&=\sqrt{q}q^{ab}\delta q_{ab}(D^aND_au)-2\sqrt{q}\delta q_{cd}D^dND^cu\,,
\end{align}
such that the variation with respect to $q_{ij}$ is obtained as
\begin{equation}
\label{vartermo2}
\frac{\delta}{\delta q_{ab}}(-2N\sqrt{q}D^cD_cu)=\frac{\sqrt{q}}{2\kappa}\bigg(q^{ab}D^cND_cu-2D^{(a}ND^{b)}u\bigg)\,.
\end{equation}
The nonalgebraic contributions in the variations for $u$ arise from the terms $\frac{N\omega}{1-u}D_cuD^cu$ and $-2N D^cD_cu$. By denoting $\delta_u$ as the variation with respect to $u$, varying the first provides
\begin{align}
\label{vartermo3}
\delta_u\bigg(\frac{N\omega}{1-u}D_cuD^cu\bigg)&=\delta_u\bigg(\frac{N\omega}{1-u}\bigg)D_cuD^cu+\frac{\omega}{1-u}\bigg(ND_c\delta uD^c u+ND_c uD^c\delta u\bigg) \notag \\
&\simeq \delta_u \bigg(\frac{N\omega}{1-u}\bigg)D_cuD^cu+\frac{\omega}{1-u}\bigg(-D_cND^cu\delta u-N\delta uD_cD^cu \notag \\
&\phantom{{}={}}\qquad -D^cND_cu\delta u-N\delta uD^cD_c u\bigg) \notag \\
&=\delta_u\bigg(\frac{N\omega}{1-u}\bigg)D_cuD^cu+\frac{2\omega}{1-u}\bigg(-D_cND^cu-ND_cD^cu\bigg)\delta u\,,
\end{align}
where we benefited from
\begin{equation}
D_c(ND^cu\delta u)=D_cND^cu\delta u+ND_cD^cu\delta u+ND^cuD_c\delta u\,.
\end{equation}
The first term in eq.~\eqref{vartermo3} as well as the double-derivative term of $u$ do not contribute to the Poisson bracket of smeared constraints, as they cancel with their symmetric counterparts when $N\leftrightarrow N'$.

Finally, the variation of $-2N D^cD_cu$ for $u$ is
\begin{equation}
\label{vartermo3.1}
\delta_u(-2N D^cD_cu)\simeq -2D_cD^cN\delta u\,.
\end{equation}
All the ingredients for the evaluation of the Poisson brackets between smeared Hamiltonian constraints, eq.~\eqref{PB_Hamiltoniano}, are at our disposal. Now,  $\delta H[N]/\delta \pi^{ab}$ and $\delta H[N]/\delta \pi_u$ are taken from eqs.~\eqref{eq:poisson-bracket-hamiltonian-constraint-q} and \eqref{eq:poisson-bracket-hamiltonian-constraint-u}, respectively. Substituting the latter and the findings of eqs.~\eqref{vartermo1}, \eqref{vartermo2}, \eqref{vartermo3}, and \eqref{vartermo3.1} into \eqref{PB_Hamiltoniano}, we arrive at:
\begin{align}
 \{H[N], H[N']\}&=\int_{\Sigma_t} \mathrm{d}^3z\bigg[\bigg(q^{ab}\Big(D^c D_c [(1-u)N]+D^cND_cu\Big)-D^a D^b [(1-u)N] \notag \\
&\phantom{{}={}}\hspace{1.3cm}-2D^{(a}ND^{b)}u\bigg)\frac{N'}{1-u}\bigg[2\bigg(\pi_{ab}-\frac{\pi}{2}q_{ab}\bigg)+\frac{q_{ab}}{3+2\omega}[(1-u)\pi_u+\pi]\bigg] \notag \\
&\phantom{{}={}}\hspace{1.3cm}-2\bigg(\frac{\omega}{1-u}D_dND^du+D_dD^dN\bigg)\frac{N'}{3+2\omega}[\pi+(1-u)\pi_u] \notag \\
&\phantom{{}={}}\hspace{1.3cm}-(N\leftrightarrow N')\bigg]\,.
\end{align}
The first line on the right-hand side of the previous equation is reformulated by using
\begin{subequations}
\begin{align}
-D^aD^b[(1-u)N]&=(D^aD^bu)N+2D^{(a}uD^{b)}N-(1-u)D^aD^bN\,, \\[1ex]
q^{ab}D^cD_c[(1-u)N]&=-q^{ab}(D^cD_cu)N-2q^{ab}D_cuD^cN+q^{ab}(1-u)D^cD_cN\,.
\end{align}
\end{subequations}
It is useful to rearrange some terms in a more suitable form:
\begin{align}
\label{algebra}
\{H[N], H[N']\}&=\int_{\Sigma_t} \mathrm{d}^3z\bigg[\bigg(-(D^a D^bN)N'+q^{ab}(D^c D_cN)N'\bigg)(2\pi_{ab}-\pi q_{ab}) \notag \\
&\phantom{{}={}}\hspace{1.3cm}+\bigg(-(D^a D^bN)N'+q^{ab}(D^c D_cN)N'\bigg)\frac{q_{ab}}{3+2\omega}[(1-u)\pi_u+\pi] \notag \\
&\phantom{{}={}}\hspace{1.3cm}-q^{ab}(D^cND_cu)\frac{N' }{1-u}\bigg[2\pi_{ab}-\frac{q_{ab}}{3+2\omega}[2(1+\omega)\pi-(1-u)\pi_u]\bigg] \notag \\
&\phantom{{}={}}\hspace{1.3cm}-2\bigg(\frac{\omega}{1-u} D_cND^cu+D_cD^cN\bigg)\frac{N'}{3+2\omega}[\pi+(1-u)\pi_u] \notag \\
&\phantom{{}={}}\hspace{1.3cm}-(N\leftrightarrow N')\bigg]\,.
\end{align}
Note that the contribution in the first line is reminiscent of a corresponding one that occurs in GR. In the following, we will be using the notation `$A\supset B$' to indicate that $B$ involves only a subset of all the terms composing $A$. We recast a part of eq.~\eqref{algebra} into
\begin{align}
\{H[N], H[N']\}&\supset\int_{\Sigma_t}\mathrm{d}^3z\,\bigg(-(D^a D^bN)N'+q^{ab}(D^c D_cN)N'\bigg)(2\pi_{ab}-\pi q_{ab})-(N\leftrightarrow N')  \notag \\
&=\int_{\Sigma_t}\mathrm{d}^3z\,2\bigg[-N'(D^a D^bN)\bigg(\pi_{ab}-\frac{\pi}{2}q_{ab}\bigg)+N'q^{ab}(D^c D_cN)\bigg(\pi_{ab}-\frac{\pi}{2}q_{ab}\bigg)\bigg] \notag \\
&\hspace{2.1cm}-(N\leftrightarrow N') \notag \\
&\simeq\int_{\Sigma_t}\mathrm{d}^3z\,2\bigg(D^bND^a\bigg[N'\bigg(\pi_{ab}-\frac{\pi}{2}q_{ab}\bigg)\bigg]-\frac{N'\pi}{2}D^cD_cN\bigg)-(N\leftrightarrow N') \notag \\
&=\int_{\Sigma_t}\mathrm{d}^3z\,2\bigg[D^bND^aN'\bigg(\pi_{ab}-\frac{\pi}{2}q_{ab}\bigg)+N'D^bND^a\bigg(\pi_{ab}-\frac{\pi}{2}q_{ab}\bigg) \notag \\
&\hspace{2.1cm}+\frac{1}{2}(D_cND^c)(N'\pi)\bigg]-(N\leftrightarrow N') \notag \\
&=\int_{\Sigma_t}\mathrm{d}^3z\,2\bigg[D^bND^aN'\bigg(\pi_{ab}-\frac{\pi}{2}q_{ab}\bigg)+N'D^bND^a\bigg(\pi_{ab}-\frac{\pi}{2}q_{ab}\bigg)\bigg] \notag \\
&\phantom{{}={}}\hspace{1.6cm}+D_cND^cN'\pi+D_cND^c\pi N'-(N\leftrightarrow N')\,.
\end{align}
Terms with single derivatives acting on $N$ and $N'$, respectively, cancel with their counterparts under $N\leftrightarrow N'$. Then, are left with
\begin{align}
\label{part.1}
\{H[N], H[N']\}&\supset\int_{\Sigma_t}\mathrm{d}^3z\,\bigg[2N'D^bND^a\bigg(\pi_{ab}-\frac{\pi}{2}q_{ab}\bigg)+N'D_cND^c\pi\bigg]-(N\leftrightarrow N') \notag \\
&=\int_{\Sigma_t}\mathrm{d}^3z\,2N'D^bND^a(q_{ac}q_{bd}\pi^{cd})-(N\leftrightarrow N') \notag \\
&=\int_{\Sigma_t}\mathrm{d}^3z\,2N'(D^bN)q_{bd}(D_c\pi^{cd})-(N\leftrightarrow N') \notag \\
&=\int_{\Sigma_t}\mathrm{d}^3z\,N(D^bN')-N'(D^bN)](-2D_c\pi^{c}_{\phantom{c}b})\,.
\end{align}
Let us now simplify the contributions in eq.~\eqref{algebra} that involve double-derivatives of $N$:
\begin{align}
\label{part.2}
\{H[N], H[N']\}&\supset\int_{\Sigma_t}\mathrm{d}^3z\,\bigg[\bigg(-(D^a D^bN)N'+q^{ab}(D^c D_cN)N'\bigg)\frac{q_{ab}}{3+2\omega}[(1-u)\pi_u+\pi] \notag \\
&\phantom{{}={}}\hspace{1.4cm}-2(D^cD_cN)\frac{N'}{3+2\omega}[\pi+(1-u)\pi_u]\bigg]-(N\leftrightarrow N') \notag \\
&=\int_{\Sigma_t}\mathrm{d}^3z\,\bigg[\bigg(-(D^c D_cN)N'+3(D^c D_cN)N'\bigg)\frac{N'}{3+2\omega}[(1-u)\pi_u+\pi] \notag \\
&\phantom{{}={}}\hspace{1.4cm}-2(D^cD_cN)\frac{N'}{3+2\omega}[\pi+(1-u)\pi_u]\bigg]-(N\leftrightarrow N') \notag \\
&=\int_{\Sigma_t}\mathrm{d}^3z\,\bigg[2(D^cD_cN)\frac{N'}{3+2\omega}[\pi+(1-u)\pi_u] \notag \\
&\hspace{1.9cm}-2(D^cD_cN)\frac{N'}{3+2\omega}[\pi+(1-u)\pi_u]\bigg]=0\,.
\end{align}
Finally, we consider the remaining terms in eq.~\eqref{algebra}:
\begin{align}
\label{part.3}
\{H[N], H[N']\}&\supset\int_{\Sigma_t}\mathrm{d}^3z\,\frac{N' }{1-u}(-q^{ab}D^cND_cu)\bigg(2\pi_{ab}-\frac{q_{ab}}{3+2\omega}[2(1+\omega)\pi-(1-u)\pi_u]\bigg) \notag \\
&\hspace{1.7cm}-\frac{N'}{1-u}\frac{2\omega }{3+2\omega}[\pi+(1-u)\pi_u]D^cND_cu-(N\leftrightarrow N')
\notag \\
&=\int_{\Sigma_t}\mathrm{d}^3z\,\frac{N' }{1-u}\bigg(-2\pi+\frac{3}{3+2\omega}[2(1+\omega)\pi-(1-u)\pi_u]\bigg)D^cND_cu \notag \\
&\hspace{1.7cm}- \frac{N'}{1-u}\frac{2\omega}{3+2\omega}[\pi+(1-u)\pi_u]D^cND_cu-(N\leftrightarrow N') \notag \\
&=\int_{\Sigma_t}\mathrm{d}^3z\,\frac{N' }{1-u}\bigg(\frac{2\omega\pi-3(1-u)\pi_u}{3+2\omega}\bigg)D^cND_cu \notag \\
&\hspace{1.7cm}-\frac{N' }{1-u}\frac{2\omega}{3+2\omega}[\pi+(1-u)\pi_u]D^cND_cu-(N\leftrightarrow N') \notag \\
&=\int_{\Sigma_t}\mathrm{d}^3z\,\frac{N' }{1-u}\bigg(\frac{2\omega\pi-3(1-u)\pi_u-2\omega\pi-2\omega(1-u)\pi_u}{3+2\omega}\bigg)D^cND_cu \notag \\
&\hspace{1.7cm}-(N\leftrightarrow N') \notag \\
&=\int_{\Sigma_t}\mathrm{d}^3z\,\frac{N'}{1-u}\bigg(\frac{-(3+2\omega)(1-u)\pi_u}{3+2\omega}\bigg)D^cND_cu-(N\leftrightarrow N') \notag \\
&=\int_{\Sigma_t}\mathrm{d}^3z\,(-N'D^cN)D_cu\pi_u-(N\leftrightarrow N') \notag \\
&=\int_{\Sigma_t}\mathrm{d}^3z\,[ND^cN'-N'D^cN]D_cu\pi_u\,.
\end{align}
Therefore, summing eqs.~\eqref{part.1}, \eqref{part.2}, and \eqref{part.3} and inserting the latter into eq.~\eqref{algebra}, we conclude that
\begin{equation}
\label{eq:hamiltonian-hamiltonian-smeared}
\{H[N], H[N']\}=\int_{\Sigma_t} \mathrm{d}^3z\,[ND^cN'-N'D^cN]\mathcal{H}_c=P[N\mathbf{D}N'-N'\mathbf{D}N]\,,
\end{equation}
with the momentum constraints of eq.~\eqref{eq:momentum-constraint}. So this provides eq.~\eqref{eq:constraint-algebra-2}.

\subsection{Hamiltonian with momentum constraint}
\label{sec:hamiltonian-momentum-constraint}

The mixed Poisson bracket can be computed either by brute force or a trick \cite{Jha:2022svf} taking advantage from eqs.~\eqref{eq:poisson-brackets-momentum-constraint}. We follow the second approach, since it is more transparent and cast the mixed Poisson bracket into the following form:
\begin{align*}
\label{eq:mixed-Poisson}
\{P[\mathbf{N}], H[N]\}&=\int_{\Sigma_t} \mathrm{d}^3z\,\bigg[\frac{\delta P[\mathbf{N}]}{\delta q_{ab}}\frac{\delta H[N]}{\delta \pi^{ab}}-\frac{\delta H[N]}{\delta q_{ab}}\frac{\delta P[\mathbf{N}]}{\delta \pi^{ab}}+\frac{\delta P[\mathbf{N}]}{\delta u}\frac{\delta H[N]}{\delta \pi_{u}}-\frac{\delta H[N]}{\delta u}\frac{\delta P[\mathbf{N}]}{\delta \pi_{u}}\bigg] \\
&=-\int_{\Sigma_t} \mathrm{d}^3z\bigg[\mathcal{L}_{\textbf{N}}\pi^{ab}\frac{\delta H[N]}{\delta \pi^{ab}}+\mathcal{L}_{\textbf{N}}q_{ab}\frac{\delta H[N]}{\delta q_{ab}}+\mathcal{L}_{\textbf{N}}\pi_{u}\frac{\delta H[N]}{\delta \pi_{u}}+\mathcal{L}_{\textbf{N}}u\frac{\delta H[N]}{\delta u}\bigg]\,.
\end{align*}
Recalling the definition of the smeared Hamiltonian constraint of eq.~\eqref{eq:hamiltonian-constraint-smeared}, we can say that
\begin{equation}
H[\mathcal{L}_{\mathbf{N}}N]=\int_{\Sigma_t}\mathrm{d}^3z\,(\mathcal{L}_{\mathbf{N}}N)\mathcal{H}\,.
\end{equation}
Now, by using the chain rule,
\begin{equation}
\mathcal{L}_{\mathbf{N}}\mathcal{H}(q_{ab},u,\pi^{ab},\pi_u)=\frac{\delta\mathcal{H}}{\delta\pi^{ab}}\mathcal{L}_{\mathbf{N}}\pi^{ab}+\frac{\delta\mathcal{H}}{\delta q_{ab}}\mathcal{L}_{\mathbf{N}}q_{ab}+\frac{\delta\mathcal{H}}{\delta\pi_u}\mathcal{L}_{\mathbf{N}}\pi_u+\frac{\delta\mathcal{H}}{\delta u}\mathcal{L}_{\mathbf{N}}u\,,
\end{equation}
which allows us to infer that
\begin{equation}
\{P[\mathbf{N}],H[N]\}=-\int_{\Sigma_t}\mathrm{d}^3z\,N\mathcal{L}_{\mathbf{N}}\mathcal{H}\simeq \int_{\Sigma_t}\mathrm{d}^3z\,(\mathcal{L}_{\mathbf{N}}N)\mathcal{H}\,,
\end{equation}
where we discarded a surface term. Therefore, we can immediately conclude that
\begin{equation}
\label{eq:momentum-hamiltonian-smeared}
\{P[\mathbf{N}],H[N]\}=H[\mathcal{L}_{\mathbf{N}}N]=H[N^cD_cN]\,,
\end{equation}
which corresponds to eq.~\eqref{eq:constraint-algebra-3}. For this derivation to work, the validity of eqs.~\eqref{eq:poisson-brackets-momentum-constraint} is paramount. The latter is guaranteed, since spatial diffeomorphisms are broken spontaneously.

\subsection{Removal of smearing functions}
\label{sec:removal-smearing-functions}

The Poisson brackets between Hamiltonian and momentum constraints are sometimes provided without resorting to smearing functions. The smearing can be gotten rid of in replacing the smearing functions by suitably chosen Kronecker symbols and Dirac functions as follows:
\begin{align*}
P[\mathbf{N}]&=\int_{\Sigma_t}\mathrm{d}^3z N^a(z)\mathcal{H}_a(z)\rightarrow \int_{\Sigma_t}\mathrm{d}^3z\,\delta^a_{\phantom{a}b}\delta(z-x')H_a(z)=H_b(x') \\[1ex]
H[N]&=\int_{\Sigma_t}\mathrm{d}^3z\,N(z)\mathcal{H}(z)\rightarrow \int_{\Sigma_t}\mathrm{d}^3z\,\delta(z-x)\mathcal{H}(z)=\mathcal{H}(x)\,.
\end{align*}
Note that the left-hand and right-hand sides are not equal to each other, but the correspondences are expressed by arrows. Then, evaluating eq.~\eqref{eq:momentum-momentum-smeared} in the same manner leads to
\begin{align*}
P[(\mathbf{N}\cdot\mathbf{D})\mathbf{N}'-(\mathbf{N}'\cdot\mathbf{D})\mathbf{N}]&=\int_{\Sigma_t}\mathrm{d}^3z\,\Big[N^b(z)D_b(z)N'^a(z)-N'^b(z)D_b(z)N^a(z)\Big]\mathcal{H}_a(z) \notag \\
&\rightarrow \int_{\Sigma_t}\mathrm{d}^3z\,\Big[\delta^b_{\phantom{b}c}\delta(z-x')D_b\delta^a_{\phantom{a}d}\delta(z-x'')\mathcal{H}_a(z) \notag \\
&\hspace{2cm}-\delta^b_{\phantom{b}d}\delta(z-x'')D_b\delta^a_{\phantom{a}c}\delta(z-x')\mathcal{H}_a(z)\Big] \notag \\
&=\mathcal{H}_d(x')D_c\delta(x'-x'')-\mathcal{H}_c(x'')D_d\delta(x''-x')\,.
\end{align*}
Furthermore, eq.~\eqref{eq:hamiltonian-hamiltonian-smeared} can be reformulated as follows:
\begin{align*}
P[N\mathbf{D}N'-N'\mathbf{D}N]&=\int_{\Sigma_t} \mathrm{d}^3z\,[ND^aN'-N'D^aN]\mathcal{H}_a \notag \\
&\rightarrow\int_{\Sigma_t}\mathrm{d}^3z\,[\delta(z-x')D^a\delta(z-x'')-\delta(z-x'')D^aN(z-x')]\mathcal{H}_a(z) \notag \\
&=\mathcal{H}_a(x')D^a\delta(x'-x'')-\mathcal{H}_a(x'')D^a\delta(x''-x')\,.
\end{align*}
Finally, eq.~\eqref{eq:momentum-hamiltonian-smeared} provides
\begin{align*}
H[(\mathbf{N}\cdot\mathbf{D})N]&=\int_{\Sigma_t}\mathrm{d}^3z\,N^a(z)D_aN(z)\mathcal{H}(z) \notag \\
&\rightarrow\int_{\Sigma_t}\mathrm{d}^3z\,\delta^a_{\phantom{a}b}\delta(z-x')D_a\delta(z-x'')\mathcal{H}(z)=\mathcal{H}(x')D_b\delta(x'-x'')\,.
\end{align*}
Since the covariant derivatives now act on scalar functions, they can be replaced by partial derivatives. The latter results are found in eq.~\eqref{eq:constraint-algebra-without-smearing} of the main text.






\bibliographystyle{JHEP}

\end{document}